\documentstyle[aps,pra,epsf]{revtex} 
\begin{document}
\draft
\tighten
\twocolumn[\hsize\textwidth\columnwidth\hsize\csname
@twocolumnfalse\endcsname
\title{Hilbert space structure of a solid state 
quantum computer: two-electron states of a double quantum dot 
artificial molecule}
\author{Xuedong Hu and S. Das Sarma}
\address{Department of Physics, University of Maryland, College
Park, MD 20742-4111}
\date{\today}
\maketitle
\begin{abstract}
We study theoretically a double quantum dot hydrogen 
molecule in the GaAs conduction band as the
basic elementary gate for a quantum computer with the electron
spins in the dots serving as qubits.  Such a two-dot 
system provides the
necessary two-qubit entanglement required for quantum
computation.  We determine the excitation spectrum of
two horizontally coupled quantum dots with two
confined electrons, and study its
dependence on an external magnetic field.  In particular,
we focus on the splitting of the lowest singlet and triplet 
states, the double occupation probability of the lowest states,
and the relative energy scales of these states.
We point out that at zero magnetic field it is difficult to
have both a vanishing double occupation probability for a
small error rate and a sizable exchange coupling for fast
gating.  On the other hand, finite magnetic fields may 
provide finite exchange coupling for quantum computer operations
with small errors.
We critically discuss the applicability of the envelope function
approach in the current scheme and also the merits of
various quantum chemical approaches
in dealing with few-electron problems in quantum dots,
such as the Hartree-Fock self-consistent field method, 
the molecular
orbital method, the Heisenberg model, 
and the Hubbard model. 
We also discuss a number of relevant issues in
quantum dot quantum computing in the context of our calculations, 
such as the required design tolerance, 
spin decoherence, adiabatic
transitions, magnetic field control, and error correction. 
\end{abstract}

\vskip2pc]
\narrowtext

\section{Background}

In recent years there has been a great deal of (as well as 
a growing) interest throughout the physics community in 
quantum computation and quantum computers (QC) \cite{Reviews}, 
in which microscopic degrees of freedom such as atomic 
levels and electron spins play the role of quantum bits (qubits).  
Because of the inherent entanglement and superposition during 
the {\em unitary} evolution of multiple qubits, 
QCs can perform certain tasks such
as factoring large integers \cite{Shor} exponentially faster than
classical computers.  They also have significant advantages over
classical computers in tasks such as searching \cite{Grover} and
simulating quantum mechanical systems \cite{Feynman,Cory}.  
Moreover, quantum error correction codes have been discovered
\cite{error}, which further bolster the hope for a practical
quantum computer.
Various QC architectures have been proposed in the literature.
The basic ingredients for a QC are two-level 
elements serving
as qubits, controlled single- and two-qubit unitary operations, 
exponentially large and precisely defined (i.e., no mixing
with other states) Hilbert space, weak
decoherence, and single qubit measurements
\cite{DiVincenzo}.  
One of the earliest QC proposals uses electronic energy
levels of ions in a linear trap (``ion trap'' QC)
as qubits \cite{Zoller,iontrap}.  
Optical pulses perform single-qubit operations,
while two-qubit operations are provided by multiple optical
pulses with the lowest vibrational mode
of the ion chain as an intermediary.  In another proposed
QC architecture, the 
cavity QED QC 
\cite{Sleator,cavityQED}, photon polarization provides 
the two required states for a qubit.  
The polarization state can be rotated optically, which provides
single-qubit operations.  Two-qubit operations are achieved 
with the intermediary of a trapped atom in the cavity
using the atom-photon interaction.  
Yet another proposed QC architecture uses 
bulk NMR techniques, where the individual nuclear spins in a 
molecule are the qubits, with different locations 
on the molecule as tags for each qubit.
Radio frequency electromagnetic pulses provide single-qubit
operations while dipolar interaction between nuclear spins
is used for two-qubit operations.  Final state 
detection in the NMR QC is achieved through an 
ensemble average over
all the molecules in the entire bulk solution.
Single- and two-qubit operations have so far
been demonstrated in trapped ions \cite{iontrap},
photons in a microcavity \cite{cavityQED}, and nuclear spins
in bulk solutions \cite{NMR}.  One perceived shortcoming of all 
these approaches, however, is their lack of scalability.
For example, it has been pointed out that the NMR approach
(which is considered to be a promising QC architecture)
cannot go beyond 20 qubits because of exponentially
diminishing signal to noise ratio as the number of
qubits increases \cite{NMR}.  
For the ion trap and cavity QED QC systems it is hard 
to see how one would surpass only a few qubits.  Thus the
atomic/molecular systems, which have so far demonstrated 
single- and perhaps even two-qubit operations, are unlikely to 
lead to an operational QC due to severe scalability problems.

There have been several recent proposals for solid state 
quantum computers, with superconducting Cooper pairs
\cite{Schon}, electron spins \cite{LD,Imam}, electron orbital
energy levels in nanostructures \cite{Steel}, 
and donor nuclear spins \cite{BKane1} serving
as qubits.  A solid state quantum computer, if it can ever be 
built, holds 
a decisive advantage in scalability compared to the
atomic/molecular systems mentioned above.  
However,
strong decoherence, intrinsic difficulty in obtaining
precise microscopic control, 
and inherently complicated Hilbert space are 
key roadblocks on the way to a practical solid state QC.
In fact, no demonstration of even a single-qubit operation
has yet been achieved in a solid state QC, and thus the
subject of developing reasonable QC hardwares faces
the unenviable dichotomy: QC systems with demonstrated
qubits are difficult to scale up, while proposed scalable QC 
systems do not have any qubits.

In this paper we study the solid state QC architecture
which uses electron 
spins in two-dimensional horizontally coupled quantum dots as 
qubits \cite{LD}.  Here a single electron is trapped in each 
individual quantum dot.  Spins of these trapped electrons are 
qubits, while the quantum dots in which they reside provide 
tags for each qubit.  Single-qubit operations,
involving the modification of local electronic spin
states in each dot, are to be 
performed using external local magnetic
field pulses, while two-qubit operations are realized using
the exchange interaction between two electrons in neighboring
quantum dots. 
Since electron spin eigenstates usually have very long 
coherence times
compared with electron orbital states \cite{Fabian1,Fabian2}, 
spin states may be better candidates for the role of qubits.  
Although a microscopically local
magnetic field is not a standard feature of modern condensed matter
experiments, reasonable proposals for the local manipulation
of spin states have been put forward \cite{LD}.  
Exchange interaction can be tuned by various means, including 
external gate potentials and external magnetic fields.   
An important point is that a single electron spin can, in
principle, be detected by SQUID magnetometers, and
it has been proposed that single electron 
spin detection can also be done by transferring the spin information
to charge degrees of freedom, which can then be detected
via the sensitive single electron transistor 
technique \cite{BKane2}.
The spin-based quantum dot quantum computer
proposal clearly has important merits and deserves serious
consideration.  Much theoretical work, however, is needed
to investigate whether the design tolerance 
required for QC operations can actually be achieved
in the state of the art quantum dot systems.  In this
study we focus on the Hilbert space structure of
coupled quantum dot systems and its implications for 
quantum computing.  One of our specific goals is to ascertain,
through fairly extensive numerical computations, whether
the spin-based quantum dot QC is a feasible proposal
even from an idealized theoretical perspective.  We
believe that such a theoretical study is necessary before
one could seriously consider the fabrication of a 
quantum dot QC.

\section{Introduction}

There are many different ways to fabricate a quantum dot
\cite{Jacak}.  In GaAs, which is the system we nominally
consider in this paper,
one common approach is
to apply external electric fields through lithographically 
patterned gates to produce a depletion
area in a 2-dimension electron gas (2DEG).  In particular, 
nanoscale electrodes are created on the surface of a 
heterostructure using photo-lithography.  The application of
appropriate electric voltages over the electrodes then produces
a suitable confining potential, thus
creating areas where electrons have been pushed away 
at desired locations (depletion areas). The typical size 
of this type of dots with the currently available 
lithographic techniques is
generally large (in the order of $100 \times 100$ to
$1000 \times 1000$ nm$^2$).  The important physical parameters
for such a quantum dot are the shape and strength of the
confinement potential, number of electrons trapped, the strength 
of the
electron-electron interaction, the strength of the additional 
external fields (magnetic, electric, etc.), impurities, 
surface roughness, boundary
irregularities, etc.  Substantially smaller size quantum dots 
can be made by direct material growth techniques, such as
quantum dot self-assembly, but it has been difficult to 
add electrons to such self-assembled dots, making QC
architecture difficult.

The study of semiconductor quantum dots and other 
nanostructures has been a large and 
fast developing field in the past ten years \cite{Ashoori}.
There are, however, relatively few works concentrating on 
the properties of two electrons in a coupled double dot
system, or an artificial hydrogenic quantum dot molecule,
which is the subject of our work.
Among related studies, quantum dot Helium (two electrons in a single 
quantum dot) has been theoretically investigated 
\cite{Merkt,Pfannkuche}.  A vertically coupled double-quantum-dot 
system has also been theoretically 
\cite{Palacios,Tamura,Asano,Partoens,BSL} 
and experimentally
\cite{Schmidt,Austing,Fujisawa1,Fujisawa2} studied.  
The horizontally coupled double-dot ``hydrogen'' molecule, 
which 
is the focus of the current paper, has been studied 
experimentally and theoretically in the context of
transport and optical (or infrared) spectroscopic experiments 
\cite{Waugh,Livermore,Oosterkamp,Blick,Yanno},
and very recently
the case when there are only two electrons in the double-dot
structure has been treated theoretically \cite{BLD} in a rather
simple approximation scheme 
using the Heitler-London and the Hund-Mulliken molecular 
orbital approaches.  An additional complication in the case of 
horizontally coupled dots is that the z direction angular momentum
is not conserved because of the absence of the
cylindrical symmetry, while this symmetry can be used to 
simplify calculations in the case of single quantum dot or 
a vertically coupled dot.  This lack of the z-angular momentum
conservation makes our calculation substantially more 
complicated than earlier quantum dot electronic structure
calculations 
\cite{Merkt,Pfannkuche,Palacios,Tamura,Asano,Partoens} in
single dot and vertically coupled dot structures.

In this paper we present our study of
the Hilbert space structure of a 
horizontally-coupled double quantum dot system
as shown schematically in Fig. 1.  
Such a horizontally coupled double quantum dot system,
with suitable lithographic gates to control the inter-dot
coupling, is one of the minimal requirements for a
spin-based quantum dot QC.
Vertically-coupled double dots might not be as good a candidate
for the purpose of quantum computing
because the coupling between the dots cannot be tuned as easily,
while the tuning of inter-dot coupling is essential
in the two-qubit operations.
QC operation requires a very special Hilbert space structure 
with a very large and precisely defined state space.  
In the electron-spin-as-qubit proposal we consider in this 
paper, one 
crucial condition is the isolation of the electron spins 
from their environment, including the electronic orbital 
degrees of freedom.  For example, if a doubly occupied
state (with two electrons in the same orbital state
of a single quantum dot)
is easily accessed, when the two electrons separate
again, one loses all the information about the identification 
of the spins (the ``tags'').  Therefore, one stringent 
requirement is that the Hilbert space should be such as
not to allow appreciable double occupation.
This is, however, quite tricky since the double occupation
probability obviously depends on inter-dot tunneling
which cannot be zero if there is to be an appreciable
exchange coupling (which is required for two-qubit
operations in the current model).
The goal of the current study is to
obtain the Hilbert space for a two-electron double dot system
using reasonably realistic quantum chemical techniques.
Since single-qubit and two-qubit operations are the only
operations necessary for quantum computing \cite{Reviews}, 
our study would be exploring the envelope of the needed 
Hilbert space (for QC) and its proximity to the unwanted 
excited state
space.  We are to assess the constraints and tolerance
required to fabricate a spin-based quantum dot QC system.
We will go beyond the simple Heitler-London and
Hund-Mulliken models and take into consideration electron
correlation through a bigger basis in the molecular orbital
calculation.  We use several approximations of varying 
complexity in our electronic structure calculations in 
order to obtain a realistic estimate of the theoretical 
computational work which will be needed to provide the
underlying basis for fabricating a spin-based quantum dot 
QC.

\section{Theory}
\label{sect:theory}

\subsection{Model Hamiltonian}

In the current study we use a single conduction band
effective mass envelope function to
describe the confined electrons in two dimensional (2D)
GaAs quantum dots.  
Such an approach is valid if the characteristic energy 
corresponding to the envelope function is much smaller than
the fundamental band gap.  In addition,
the excitation energy along the third (growth) direction
has to be much larger than all the characteristic 
2D excitation energies.  In the case of GaAs,
the fundamental gap is 1.5 eV.  
Furthermore, for a 10 nm thick 2D GaAs
quantum well (which hosts the quantum dot), 
the first intersubband 
excitation energy (for excitations along the growth direction)
is typically 0.1 eV.  
Since the characteristic in-plane 2D excitation 
energy of the confined electron(s) is in the order of 
1 to 10 meV, the applicability criterion for the effective
mass single envelope function approximation is well satisfied.  
The effective two-electron quantum dot
molecule Hamiltonian in the presence of an external magnetic
field (defined through the vector potential ${\bf A}$) is then
\begin{eqnarray}
H&=&\sum_{i=1}^{2} \left[ \frac{1}{2m^*} \left( {\bf p} 
+ \frac{e}{c}{\bf A}({\bf r}_i) \right)^2 + V({\bf r}_i) 
\right. \nonumber \\
& & \left. + g^* \mu_B {\bf B} 
\cdot {\bf S}_i \right] + \frac{e^2}{\epsilon r_{12}} \,,
\label{eq:Hamiltonian}
\end{eqnarray}
where $m^*$ is the conduction electron effective mass, 
$V({\bf r}_i)$ 
is the quantum dot potential (which is to be parametrized
in our work, but can in principle be calculated by
self-consistent techniques if the details of the 
electrostatic confinement are known), 
$g^*$ is the effective 
electron g-factor, $\mu_B$ is the Bohr magneton, $g^*\mu_B 
{\bf B}\cdot {\bf S}_i$ is the Zeeman splitting, $\epsilon$ 
is the static background (lattice) dielectric constant, 
and $r_{12}$ 
is the distance between the two electrons.  
Here we uncritically assume the effective mass approximation
(which we will show to be well-valid)
assuming the interband mixing to be negligibly 
small in the low energy sector of our interest and the
effect of the periodic crystal potential to be described by
the electron effective mass and the background dielectric 
constant.  
The quantum well material
we focus on in this work is GaAs, 
thus $g^* \approx -0.44$, $\epsilon \approx 13.1$, and $m^*
\approx 0.067m_0$, where $m_0$ is the bare electron mass.
We use as our model 2D quantum dot
confinement potential the following linear combination 
of three Gaussians defined by the adjustable parameters 
$V_0$, $a$, $V_b$, $l_x$, $l_y$, $l_{bx}$, and $l_{by}$: 
\begin{eqnarray}
V({\bf r})&=&V_0 \left[ \exp\left(-\frac{(x-a)^2}{l_x^2}
\right)+\exp\left(-\frac{(x+a)^2}{l_x^2}\right) \right]
 \nonumber \\
& & \hspace*{-0.2in} 
\times \exp\left(-\frac{y^2}{l_y^2}\right) +V_b
\exp\left(-\frac{x^2}{l_{bx}^2}\right)
\exp\left(-\frac{y^2}{l_{by}^2}\right) \;;
\label{eq:confinement}
\end{eqnarray}  
Here the first two Gaussians (with a strength of $V_0$)
are for the individual dot potential wells and the third 
(with a strength of $V_b$) is for 
controlling the
central barrier (so that we can adjust the barrier easily
and independent of the locations of the other two Gaussians).
Thus $V_0$ is the potential well depth while $V_b$ controls 
the central
potential barrier height.  We choose this form for the 
confinement potential mainly because of its 
simplicity and versatility, and no other particular 
significance should be attached to our choice.
To find a realistic form for $V$ requires a self-consistent
calculation using the correct boundary conditions and 
heterostructure parameters, which is not warranted at the
current level of QC modeling (and is well beyond the
scope of this work).  We only note here that the confinement
potential defined by Eq.(\ref{eq:confinement}) is a reasonable
potential for 2D double quantum dot structures defined
electrostatically provided the confinement along the growth
(z) direction is much tighter than the 2D confinement as
discussed above.  It is easy to fit a realistic confinement
potential, if available, to this simple Gaussian form.  

The two-electron Hamiltonian cannot be solved exactly.  We use 
two different approaches to calculate the approximate energy
spectra and electron states of the Hamiltonian $H$ 
defined by Eq.~(\ref{eq:Hamiltonian}).  
The first is a Hartree-Fock (HF)
calculation, where the two electrons are treated as independent
particles moving in a HF self-consistent field \cite{McWeeny}.  
The second is the so-called molecular 
orbital method, in which we use single-harmonic-well 
single-electron wavefunctions to form two-electron orbitals 
and use them as basis states to solve the Schr\"{o}dinger equation 
for the two electrons \cite{McWeeny}.  We note
that the presence of the external magnetic field makes our 
problem somewhat different from the standard quantum chemistry
calculations.

\subsection{Hartree-Fock Approximation}

In the HF approximation, an electron is moving in the 
mean field produced by all other electrons.  The
multiple-electron wavefunction is a single Slater determinant.
Pauli principle is thus obeyed so that electrons with the same spin 
do not occupy the same orbital state simultaneously.  Electron
correlation is therefore taken into account crudely
in the sense that only the Pauli-principle-imposed
correlations are included.  
There are a variety of Hartree-Fock calculations
in the context of quantum chemistry \cite{McWeeny}.  In our study
here, a restricted Hartree-Fock (RHF) calculation, where the two
electrons with different spins occupy the same spatial orbital,
significantly over-estimates the double occupation probability
and thus over-estimate the energy of a singlet state. 
Such an RHF calculation is clearly a rather poor 
approximation for our purpose where an accurate knowledge
of the double occupation probability and of the singlet-triplet
splitting is an important requirement.
Therefore, we adapt an unrestricted Hartree-Fock (UHF) approach 
\cite{Yanno},
where the two electrons in the ground state with opposite spins
are not required to occupy the same spatial orbital.  This method
inherently incorporates the uncorrelated nature of two remote 
quantum dots, which is partially satisfactory for our purpose.
However, the
shortcoming of this approach is that the ground state (where the
spins of the two electrons are opposite) is not a pure singlet
state.

The HF equations for N electrons are: 
\begin{eqnarray} 
F \psi_i({\bf r}_1) & = & E_i \psi_i({\bf r}_1)\,,  \\
F & = & f + \sum_{j=1}^{N} (J_j-K_j)\,, \nonumber \\
f & = & \frac{1}{2m^*} \left[ {\bf p} + \frac{e}{c} 
{\bf A}({\bf r}_1) \right]^2 + V({\bf r}_1)\,, \nonumber \\
J_j \psi_i ({\bf r}_1) & = & \int \psi_j({\bf r}_2)^* 
\psi_j({\bf r}_2) \; \frac{e^2}{\epsilon r_{12}} \;
d{\bf r}_2 \cdot \psi_i ({\bf r}_1)\,,
\nonumber \\
K_j \psi_i ({\bf r}_1) & = & \int \psi_j({\bf r}_2)^*
\psi_i ({\bf r}_2) \; \frac{e^2}{\epsilon r_{12}} \;
d{\bf r}_2 \cdot \psi_j ({\bf r}_1) \,. \nonumber
\end{eqnarray}
Here $\psi_i$ ($i=1 \ldots N$) are the appropriate single particle 
wavefunctions; $f$ is the single particle part of the
Fock operator $F$, which has the same form as the corresponding
terms of Eq.(\ref{eq:Hamiltonian}); the operator
$J_i$ is the direct Coulomb repulsion between two electrons;
while the operator $K_i$ is the exchange interaction between 
electrons.
All the integrals include sum over different spin indices.
In other words, the exchange term $K_j$ vanishes if the
spin indices of the $j$th and $i$th electron orbitals are
different.

The advantage of HF approximation lies in its clear
physical picture of an effective single particle 
dynamics in a self-consistent field
background.  However, its shortcoming is also because of this
simplicity: electrons only see an average background produced 
by the other charges, not the instantaneous 
locations of those charges, and therefore electron
correlation is not taken into account beyond the Pauli
principle (which is built into the Slater determinant).  
In addition, in numerically solving the HF
equations in the presence of a finite magnetic field, 
the choice of gauge turns out to be important, a fact
which we have not found to have been discussed earlier
in the literature.  When the two
electrons are well separated and each is confined to its own
well, they should have their own gauges; if the electron
wavefunctions are extended throughout the two wells, 
then a single gauge has to be used.
The use of a single gauge, however, significantly raises the
Coulomb repulsion energy of the electrons, because the ${\bf
A}^2$ term can be quite large in high magnetic fields, thus
behaving like an additional confining harmonic potential, 
pushing the two
electrons towards each other.  It is interesting to 
reflect on why gauge invariance breaks down in this 
HF calculation.  It certainly should hold for 
the exact two-electron Schr\"{o}dinger equation,
in contrast to the HF approximation. 
However, as we make the Hartree-Fock approximations, we
mostly neglect the electron-electron correlations.  The choice 
of gauge thus becomes relevant to our approximate calculation.
This point is further illustrated by the fact that the exact 
two-electron wavefunction is a superposition of an infinite
number of Slater determinants.  As these determinants
generally transform in different ways under a gauge 
transformation, 
the change in the overall wavefunction can be quite 
different from that of each individual Slater determinant.
The breakdown of the gauge invariance in the HF approximation
thus arises from its very special single Slater determinant form.
The broken gauge invariance shows a glaring weakness of the
Hartree-Fock approach which clearly needs to be supplemented 
by other techniques in order to obtain a 
more complete description of the two-electron system.  
Below we discuss another quantum chemical approach that can 
better describe the electron correlations.

\subsection{Molecular Orbital Method}

For a two-electron problem, the molecular orbital
approach involves choosing suitable single-electron 
basis functions, forming two-electron 
orbitals from these basis functions, expanding 
the two-electron Schr\"{o}dinger equation in these two-electron
orbitals, and finally solving the eigenvalue problem
(presumably through a direct matrix diagonalization).
In our molecular orbital calculation, we use the single-dot 
single-electron wavefunctions as the basis states to form 
our molecular orbitals.  These single-electron wavefunctions 
are the usual Fock-Darwin states (assuming
parabolic confinement at the bottom of the potential wells)
\cite{Jacak}.  We take care in ensuring that our two-electron
wavefunctions have the correct symmetry of our two-particle
Hamiltonian defined in Eq.~(\ref{eq:Hamiltonian}).  In the
simplest case (the so-called Hund-Mulliken approximation), 
we use only the two single-dot ground eigenstates (s-orbitals) 
as the basis states.  
These wavefunctions take the following form 
for a symmetric quantum dot structure with identical
confinement along x and y directions \cite{Jacak,BLD}
\begin{eqnarray}
\phi_{\mp a, \rm{or} \ L/R} (x,y) 
& = & \frac{1}{\sqrt{\pi} l_0} 
\exp \left[{\frac{(x \pm a)^2+y^2}{2 l_0^2}} \right] 
\nonumber \\
& & \times \exp \left(\mp i \frac{ay}{2 l_B^2} \right) \;,
\end{eqnarray}
where
\begin{eqnarray}
l_0 & = & \frac{l_B}{\sqrt[4]{1/4 + \omega_0^2/\omega_c^2}} \,,
\nonumber \\
l_B & = & \sqrt{\frac{\hbar c}{e B}} \,, \\
\omega_c & = & \frac{e B}{m^* c} \,. \nonumber 
\end{eqnarray}
Here $\pm a$ are the potential minima locations of the two
quantum dots which are horizontally placed along the 
x direction; 
$l_0$ is the effective wavefunction radius;
$l_B$ is the magnetic length for the applied magnetic field
$B$ along the z direction; 
$\omega_0$ is the confinement parabolic well frequency;
$\omega_c$ is the electron cyclotron frequency; and $m^*$
is the GaAs conduction electron effective mass.  The
gauge that produces the above wavefunction is ${\bf A}
= \frac{B}{2}(-y,x,0)$.  Note that in choosing our 
single-particle basis to form the molecular orbitals we use
the exact one-electron eigenstates corresponding to
a double parabolic well 2D potential which is obtained 
by expanding the Gaussian potential well of 
Eq.~(\ref{eq:confinement}) around its minima.  This
particular basis has the great advantage of being
analytic (the Fock-Darwin levels) as well as a
reasonable basis for the problem we are to solve.
Using the single-dot wave functions $\phi_{L/R}({\bf r})$, 
the corresponding triplet wavefunction
is $\Psi_1=[\phi_L({\bf r}_1)\,\phi_R({\bf r}_2) -
\phi_L({\bf r}_2)\,\phi_R({\bf r}_1)]/\sqrt{2}$, 
while the singlet wavefunctions are 
$\Psi_2 = [\phi_L({\bf r}_1)\,\phi_R({\bf r}_2) + 
\phi_L({\bf r}_2)\,\phi_R({\bf r}_1)]/\sqrt{2}$,  
$\Psi_3=\phi_L({\bf r}_1)\,\phi_L({\bf r}_2)$, 
and $\Psi_4=\phi_R({\bf r}_1)\,\phi_R({\bf r}_2)$.  
It is clear that this basis 
consists of the Heitler-London states $\Psi_1$ and $\Psi_2$
and the two ``ionized'' or ``polarized''
doubly-occupied states $\Psi_3$ and $\Psi_4$.  
We can solve the Schr\"{o}dinger
equation of the two-electron Hamiltonian in this basis 
by expanding in
these four functions.  Since the triplet state is antisymmetric
in the orbital degrees of freedom while singlet states are 
symmetric, they are not coupled by the symmetric Hamiltonian
of Eq.~(\ref{eq:Hamiltonian}).
Thus triplet and singlet states can be treated 
separately.  Notice that the two-electron states are generally
neither orthogonal nor normalized because the single-dot 
single-electron wavefunctions $\phi_L({\bf r})$ and 
$\phi_R({\bf r})$
are not orthogonal to each other.  Thus, the Schr\"{o}dinger
equation of the problem can be expressed as  
\begin{eqnarray}
\sum_j^4 H_{ij} c_j & = & E_i \sum_j^4 S_{ij} c_j \,,
\\
H_{ij} & = & \int \Psi^*_i(1,2) H \Psi_j(1,2) 
d{\bf r}_1 d{\bf r}_2 \,, \nonumber \\
S_{ij} & = & \int \Psi^*_i(1,2) \Psi_j(1,2) 
d{\bf r}_1 d{\bf r}_2 \,, \nonumber
\end{eqnarray}
We now have a generalized eigenvalue problem.  It can be readily
solved numerically.  Formally, it is identical to an ordinary
eigenvalue problem if we know the inverse of $S$.

To systematically improve upon the four-state molecular orbital
calculation, we include the first excited states of single
quantum dots (p-orbitals) in an improved (so-called s-p
hybridized) molecular orbital calculation.  The single
particle p-orbitals have the following forms
\begin{eqnarray}
\phi_{1,\pm 1, -a} (x,y) & = &
\frac{1}{\sqrt{\pi} l_0^2} \; [(x+a) \pm iy] \;
\exp \left[ {\frac{(x + a)^2+y^2}{2 l_0^2}} \right] 
\nonumber \\
& & \times \exp \left(-i \frac{ay}{2 l_B^2} \right) \;, \\
\phi_{1,\pm 1, a} (x,y) & = &
\frac{1}{\sqrt{\pi} l_0^2} \; [(x-a) \pm iy] \;
\exp \left[ {\frac{(x - a)^2+y^2}{2 l_0^2}} \right] 
\nonumber \\ 
& & \times \exp \left( i \frac{ay}{2 l_B^2} \right) \;. 
\end{eqnarray}
Here the first two subindices are the quantum numbers for the 
Fock-Darwin states, while the third one indicates their locations.
Now we have 6 (two s orbitals and four p orbitals)
atomic orbitals (single-electron
single-dot eigenstates), with which we can
form 21 singlet states and 15 triplet states.  Since parity
symmetry is not broken by the introduction of a magnetic field, we
can introduce even and odd single-electron molecular
orbitals, and then build the two-electron molecular orbitals
using these symmetrized orbitals.  There
are then 12 even singlet states, 9 odd singlet and triplet
states, and 6 even triplet states.  
The use of parity reduces the number of independent 
two-particle matrix elements almost by half.
The advantage of introducing the p-orbitals in the molecular 
orbital calculation is that
the excited states give us the freedom to form anisotropic
states (which could not be accomplished with the isotropic
s-orbital-only basis), 
thus enabling us to describe the electron overlap with higher
accuracy.

\section{Numerical Results}
\label{sect:results}

\subsection{Hartree-Fock Approximation}

In our HF calculation, we solve numerically the Hartree-Fock
equations by setting up a grid of $(60 \times 30)$ 
mesh points on the two-dimensional $(x, y)$ 
space.  The reason for not selecting a finer
mesh for the grid is that we have a coupled 
two-dimensional problem that is not sparse, so that the 
actual non-sparse matrix dimension reaches
$1800 \times 1800$, which is essentially our computation limit.  
We make a nonlinear transformation of the
spatial coordinates so that most of the grid points are within
the two-dot region, thus ensuring the accuracy and effectiveness
of our numerical eigensolutions.

Figs. 2 and 3 show some of the results we obtain using 
the UHF approximation.  In Fig. 2 we 
can see that the energy of the lowest 
parallel spin (triplet) state remains 
above the lowest opposite spin state and never 
dips below it up to a fairly high magnetic field of 7 Tesla
for reasonable quantum dot parameters as given in the
figure captions.  Notice that in the UHF theory 
the opposite spin state is actually a
mixture of a singlet and a triplet state, and therefore
the ground state is never a pure singlet state.  
Although an RHF
approach would have produced a pure singlet eigenstate, it 
significantly over-estimates the Coulomb energy so that the 
singlet state always has
higher energy than the triplet state, violating the theorem
that at zero magnetic field (when the wavefunction can be written 
as a real function) the ground state should be a singlet 
\cite{Ashcroft}.  In Fig. 3 we show two sets of data where the
ground-triplet splitting decreases {\it exponentially} fast 
as a function of inter-dot distance.  This suggests that,
at least in principle, an
efficient control of the splitting between the ground and 
the first excited 
state can be achieved by increasing the potential barrier
separating the dots and/or by increasing the inter-dot
separation.  

A simple Hartree-Fock calculation with a single Slater
determinant is generally not sufficiently
accurate to deal with subtle effects 
arising from small interaction terms in the Hamiltonian.  
For instance, since the reason for the
singlet-triplet crossing is essentially two-electron exchange and
correlation effects, Hartree-Fock approximation 
should not be trusted to produce quantitatively
reliable singlet-triplet splitting information (although
it is expected to be qualitatively correct since exchange,
which the HF theory includes, is expected to be the dominant
effect).  Our main reason for pursuing the HF theory, in
spite of its obvious quantitative shortcoming, is the
fact that the self-consistent HF calculation produces a more
accurate single particle wavefunction than the eigenstates of a
fixed harmonic well.  Based on these improved single particle
HF states, a configuration
interaction (CI) calculation can then be built 
in the future, which will lead
to a more faithful and quantitatively accurate 
description of the actual two-electron
wavefunctions in the double quantum dot system.  Our HF
calculation could be the starting point of such a future CI
calculation.

\subsection{Molecular Orbital Methods}

The central task in our molecular orbital calculation is
the computation of two-particle (Coulomb) matrix elements
in the molecular orbital basis set described in section 
\ref{sect:theory} of this paper.  
In the Hund-Mulliken
calculation (using only the s-orbitals)
which involves a basis of 3 
singlet states (the Heitler-London $\Psi_2$ and the two doubly 
occupied states $\Psi_3$ and $\Psi_4$) and 1 triplet state, 
we need to calculate only 7 Coulomb matrix elements (taking 
even-odd parity symmetry into consideration, only 5 
Coulomb matrix elements are needed).  
When p-orbitals are included, we need to
calculate 231 and 120 Coulomb matrix elements for the singlet and
triplet states respectively which is a substantial 
computational task.
When the even-odd symmetry is taken
into account, the number of Coulomb matrix elements
reduces to 123 and 66, respectively, for the singlet and
triplet states,
which is still a formidable task because each matrix element
corresponds to a 4-dimensional integral requiring high
accuracy.  The most computationally intensive and time consuming
part of our calculations has been the evaluation of these
Coulomb matrix elements.

Our Hund-Mulliken calculation (with only the electron s-orbitals) 
results are shown in Fig. 4.  Here we first perform a variational 
calculation at zero magnetic field.  We vary the parabolicity
and the location of the fitting parabolic well to achieve the 
lowest energy in the ground state.  The results
of the variational calculation are shown in Table 1.  
For these optimal variational
parameters the triplet state (the first excited state at zero 
and low magnetic 
field) is also quite close to its lowest energy.  
According to Fig. 4, the exchange coupling, or equivalently
the singlet-triplet 
splitting, is a sensitive function of the central barrier
height.  This implies that a suitable gate-controlled
central barrier can, in principle, be utilized to
switch on or off the exchange coupling efficiently, thereby
making possible two-qubit operations in our quantum
dot QC architecture.  The magnitude of the exchange coupling 
ranges from 0.2 meV to about 1 meV in these structures, 
which correspond to gating times in the order of one to 
tens of picosecond, which is difficult, but not
impossible, to achieve.  

The results of the molecular orbital
calculation done on the larger basis (including both single
particle s- and p-orbitals) are shown in Figs. 5-7.
Comparing this more sophisticated 
s-p hybridized calculation,
which includes the first excited ``atomic'' orbitals, 
with the simple Hund-Mulliken calculation discussed above,
we find that there is a significant effect arising from the
strong mixing-in of the higher excited states.  In other words,
s-p hybridization significantly lowers the energy of the 
lowest singlet state.  Although the s-p hybridized
ground state resemble the Heitler-London
wavefunctions, it also contains components in which one
electron is in one of the excited states.  
Such a contribution could be favorable for the quantum dot 
molecule because p-orbitals increase
the ``bonding'' between the two quantum dots, thus lowering
the overall energy of the double dot system.  
In addition,
our confinement potential is not exactly a sum of 
two symmetric
parabolic wells.  Instead, the two Gaussian wells and one
Gaussian barrier complicate the contour of confinement,
so that the true ground state has components of 
single particle excited states of the fixed harmonic well
potentials.

According to the calculated energy spectra shown in Fig. 5, 
the ground singlet and triplet states
are well separated from the rest of the 
higher excited states in the Hilbert space.  
For the representative sample parameters as chosen
the higher excited states are always separated from
the ground singlet/triplet states by at least 6 meV,  
which is much larger than the maximum value
of the exchange coupling, $J$ (0.3 meV) as well as being
much higher than $k_B T \lesssim 0.1$ meV at the
cryogenic temperature of QC operation.
Thus, as long as the coupling between the
two quantum dots is turned on slowly, the 2-spin 
two-electron system
is quite isolated from the other parts of the Hilbert space
and is thus a good candidate for a quantum gate.  This 
demonstration of a well-defined 2-spin singlet/triplet
Hilbert space, which is well-separated from the rest of the 
higher excited states of the two-electron double dot system,
is one of the most important results of our work.

Fig. 5 also shows that there are discernible shell structures 
in the two-electron excitation spectra, and this structure 
changes with the magnetic field.  The shell structure is 
especially prominent at large ${\bf B}$ field.  
The origin of the shell structure
is apparently the degeneracy of the single particle 
Fock-Darwin states.  At small ${\bf B}$ field, the 
wavefunction overlap between the two quantum dots is 
quite significant, so that the direct
Coulomb repulsion and the exchange energy
play important roles in 
deciding the energies of individual states.  As ${\bf B}$
field increases, state overlaps between two quantum dots
decreases since the wavefunctions are squeezed by the
applied magnetic field, and consequently
the Coulomb correlation between the two dots
becomes less 
important (even though the on-site Coulomb repulsion becomes
more important).  The whole spectrum should then resemble 
that of two isolated single quantum dots.
Another effect that should be taken into consideration is 
that for $|{\bf B}|>0$ the degeneracies of the Fock-Darwin
states are lifted, so the single particle energy levels are
scrambled.  However, at certain specific magnetic field
values shells appear
as several energy levels move close 
to each other and away from the rest.  
There are also apparent level crossings in the spectra, 
as the energies of individual Fock-Darwin states 
with different angular momenta change differently
with the ${\bf B}$ field, and singlet and triplet states
are not coupled by the Hamiltonian we consider.  
In summary, any simple magnetic field 
dependence of the Fock-Darwin states is scrambled by the 
non-parabolic confinement potential and the varying Coulomb
interaction between the two electrons.  
Although the origin of the shell structure is clearly 
the starting degeneracy of the Fock-Darwin levels,
its detailed magnetic field dependence is quite complex.
The shell structure may, in principle,
be useful for the purpose of quantum computing because 
a full shell plus one electron might be 
effectively considered as a spin-$\frac{1}{2}$ 
single-electron system, i.e. a
filled shell could be considered ``inert''.  
Whether such an
effective spin-$\frac{1}{2}$ system with filled shells 
is sufficient as a
qubit can only be demonstrated by a multi-electron CI
calculation of its spectrum, and clearly requires further 
investigation.  Our molecular orbital results in the
presence of the external magnetic field could be 
considered suggestive of such a possibility.

Fig. 6 shows the magnetic field dependence of the exchange 
coupling (singlet-triplet splitting) with three different 
central barrier heights.  Here
we can see that all the thick curves (from the larger basis 
calculations) are shifted upwards from the thinner curves 
(from the smaller basis calculations).  The reason for this change is 
that the larger basis allows us to obtain much lower 
(and presumably more accurate) energy
for the singlet states.  The triplet states do not
change nearly as strongly as the singlet states.  Thus the
exchange coupling $J$ changes (increases)
by 23\%, 42\%, and 18\%
respectively for 3.38, 6.28, and 9.61 meV central barriers
in the more sophisticated calculations using the larger s-p 
hybridized molecular orbital basis.
Note that the improvement in the calculated J is less for
larger central barrier potentials.  
This is consistent with our belief that the
p-orbitals play a more important role when the two-dot 
overlap is larger, and therefore s-p hybridization effects
are quantitatively more important when the inter-dot overlap
(and hence the exchange coupling) is larger.

Fig. 7 shows the ground state double occupation probability 
as a function of the magnetic field, which clearly
decreases as ${\bf B}$
field increases.  The reason for this decrease 
with increasing magnetic field is 
straightforward.  As ${\bf B}$ increases, the
single-electron atomic wavefunctions become narrower.  
Thus, the ``on-site'' Coulomb repulsion energy for the 
doubly occupied state increases
rapidly, which decreases the double occupation
probability.  The ground state double occupation probability
can also be seen in Fig. 7 to decrease significantly
with increasing central barrier strength separating
the two dots (as one would expect).
Here we do not show double occupation
for the triplet state, because in those states one electron would
have to be in an excited state, thus the probability 
is quite small, and could be considered negligible for
most purposes in contrast to the ground singlet state
situation shown in Fig. 7.
Double occupation probability is an important
parameter for a quantum dot quantum computer (QDQC).  In a
QDQC, electron spins are qubits, while their residence 
quantum dots (QD's)
(i.e. the individual dots on which the electrons are located)
are their tags to distinguish the different qubits.  If during
the gating action two electrons jump onto a single QD
due to high double occupation probability, even
if they separate eventually, their original tag information 
is lost, which will result in an error requiring appropriate
error correction.
Thus, in designing
a QDQC, one needs to minimize the double occupation probability
for the states (especially the lowest singlet state) 
that belong to the QDQC Hilbert space.  Indeed, Fig. 7 shows that
for the lower barrier cases the double occupation probabilities
are prohibitively large for the purpose of quantum computing.
On the other hand one cannot have a QDQC with very large
central barrier (thereby producing very small double 
occupation probability) because then the exchange coupling
(Fig. 6) will be very small, making 2-qubit operations
impossible.
This indicates that one has to settle for a compromise in the 
pursuit of a large exchange coupling (for achieving
smaller gating time during two-qubit operations)
and a small double occupation probability
(for reducing the error correction requirement).

To look for parameters that can lead to small double occupation
probability but still maintain a finite exchange coupling
for a double quantum dot, we increase the inter-dot distance
from 30 nm as studied above to 40 nm and perform the molecular 
orbital calculations.  The results are shown in Figs. 8-11.

Again, we first vary the location and parabolicity of the 
fitting parabolic wells.  In Fig. 8 we plot these variational
parameters as functions of the central barrier height.  One
interesting feature shown in the panel (a) of the figure is
that the fitting well parabolicity increases as the central
barrier height increases.  In other words, when the barrier 
is low, the electron wavefunctions tend to be more spread out.
Furthermore, panel (b) of Fig. 8 shows that the distance of
the two fitting wells are closer when the central barrier is
low.  These results show that the two-electron artificial
molecule is bounded tighter when the inter-dot barrier is
low, in analogy to a diatomic molecule and its orbital 
contraction.

In Figs. 9 and 10 we show the magnetic field dependence of
the exchange coupling and the double occupation probability.
The values of both these quantities at zero magnetic field
are about half of their values in Figs. 6 and 7 (with the
same barrier heights).  Therefore, at zero magnetic field,
the exchange coupling and double occupation probability 
decrease with about the same rate as we pull 
the two quantum dots 
away from each other.  In Fig. 11 we also plot the central
barrier height dependence of both the exchange coupling and 
the double occupation probability.  Both quantities decrease
exponentially as we increase the central barrier, as one
expects.  In the range of the barrier heights we considered,
the exchange coupling decreases from 0.27 meV to 0.0097 meV,
a change of about 28 times; the double occupation probability
decreases from 0.060 to 0.0017, a change of about 35 times.
Although double occupation probability decreases a little
faster than the exchange coupling, the difference is
insignificant.  Thus, we also show that these two quantities
change with about the same rate as we change the central barrier
height.  Therefore, at zero magnetic field,
it would be difficult to achieve a vanishingly small
double occupation probability while maintaining a sizable
exchange coupling.  

However, as it is shown
in Figs. 9 and 10 (and also Figs. 6 and 7), finite
magnetic field may lead to a solution to this problem
of correlated exchange coupling and double occupation probability.
Physically, the exchange coupling is determined by the
competition between the direct and exchange Coulomb 
interactions, while the double occupation probability is
mainly determined by the direct Coulomb repulsion.  It is
thus expected that the two quantities have different dependence 
on the magnetic field.  Indeed, according to Figs. 6, 7, 9, and 10,
for magnetic field above 6 Tesla, the magnitude of the exchange
coupling decreases quasi-linearly, while the double occupation
probability decreases exponentially fast.
For example, in Fig. 10, at a magnetic field of 7 Tesla and
effective central barrier of 9.61 meV, the double occupation
probability is about $6 \times 10^{-4}$, a magnitude that
is in the same order
as the tolerance of the currently available error correction 
codes; while the exchange coupling in this case is about
0.009 meV, corresponding to a swap gate time of about 3 ns
(after taking into account adiabaticity).  Thus the difference
in the magnetic field dependence of the exchange coupling and
the double occupation probability can be exploited for optimal
QC operations.  This is another important result of our calculation 
in the context of QDQC architecture.

\section{Discussions}

\subsection{Validity of the envelope function approach}

The issue of the adequacy of the single envelope function
effective mass approximation (used throughout our calculations)
for the purpose of studying electron entanglement in the
context of QDQC requires careful consideration.
Let us first discuss the validity of the envelope function
approach in our study of the electronic structure of the
2D GaAs-based double dot molecule.  
One necessary condition \cite{Bastard} 
is that the ${\bf k} \cdot {\bf p}$
approximation should be valid in our problem.  
For GaAs conduction band, the
${\bf k} \cdot {\bf p}$ approximation
(``Kane model'') is valid up to 
$\epsilon_{\bf k} - \epsilon_0 \sim 0.3$ eV, where
$\epsilon_0$ is the conduction band edge energy and 
$\epsilon_{\bf k}$ is the energy of a conduction 
electron at momentum ${\bf k}$ in the Brillouin zone.  
In our 
study of the coupled quantum dot molecule, the energy scale of the 
electrons is in the order of 10 meV, making the ${\bf k} \cdot 
{\bf p}$ approximation valid.

Another condition for the validity of the
envelope function approach
is weak inter-valley scattering.
The electronic wavefunctions in this manuscript are built 
from the 
conduction band $\Gamma$-point Bloch functions.  However, if a 
GaAs quantum well (in the growth direction we have a narrow
GaAs quantum well sandwiched between AlGaAs barriers) is 
too narrow ($<$ 3 nm), the X-valley would lie close to the 
$\Gamma$-point in energy, so that a more complete approach
(going beyond the single envelope function approach)
is needed to take into account the $\Gamma-X$
inter-valley scattering.
An envelope function approach becomes inappropriate
because it only describes {\it locally} a small part of 
the Brillouin zone.  Thus, for our
approach to be valid, the quantum well in the growth direction
cannot be too narrow \cite{Bastard}.  Calculations going beyond 
the single envelope approximation for GaAs quantum wells, 
however, show \cite{Bastard} that even for such extremely 
narrow quantum wells, the single envelope function 
approximation gives qualitatively (and semi-quantitatively)
accurate results.

Although we do not think it to be necessary at present, one 
can go beyond the single envelope function approximation.
A more complete analysis
(than our single envelope function model)
would employ an $8 \times 8$ Kane model \cite{Bastard} 
to include all the closeby valence bands, with 2 $\Gamma_6$ 
states corresponding to the
conduction bands, 4 $\Gamma_8$ states corresponding to the
heavy and light hole bands, and 2 $\Gamma_7$ states corresponding
to the split-off bands.  We would then have 8 envelope functions
instead of just one as we use here.  The complete single electron 
Schr\"{o}dinger equation
and the general QD Hamiltonian without any 
magnetic field take the forms
\begin{eqnarray}
H \psi & = & E \psi \,, \\
H & = & \frac{{\bf p}^2}{2m_0}
+ U({\bf r}) + V({\bf r}_{\perp}) \,, \nonumber \\
\psi & = & \sum_{i=1}^8 f_i u_{i0} \,. \nonumber 
\end{eqnarray}
Here $m_0$ is the bare electron mass.
$U({\bf r})$ is the crystalline periodic potential, which
assumes different values in the quantum well and in the 
barriers.  It thus has a step-like overall profile along the 
z direction.  $V({\bf r}_{\perp})$ is the QD confinement 
potential produced by an external static electric field
arising from lithographic gates, dopants, and all other
sources not contained in $U({\bf r})$.
${\bf r}_{\perp}$ refers to the 2D x-y plane, i.e.
directions perpendicular to z direction.
Since the z direction confinement is very narrow while $V$
is slowly varying, we neglect its variation along the z 
direction.  $u_{i0}$'s are the $\Gamma$ point Bloch functions,
which are the same as the atomic orbitals of the constituent
elements.  $f_i$'s are the 8 envelope functions corresponding
to the 8 relevant bands, which are slowly varying functions
on the atomic scale.  The Schr\"{o}dinger equation can be
simplified into a set of equations for the envelope functions
$f_i$'s by using the following identity
\begin{eqnarray}
\int_{\Omega} f({\bf r}) u({\bf r}) d {\bf r} 
& \cong & \frac{1}{\Omega} \int_{\Omega} u({\bf r}) d {\bf r}
\int_{\Omega} f({\bf r}) d {\bf r} \nonumber \\
& = & \frac{1}{\Omega_0} \int_{\Omega_0} u({\bf r}) d {\bf r}
\int_{\Omega} f({\bf r}) d {\bf r} \,. 
\end{eqnarray}
Here $\Omega$ is the total volume of the crystal, $\Omega_0$
is the volume of one unit cell, $f({\bf r})$ is a slowly varying 
function on the atomic scale, while $u({\bf r})$ is a fast
varying and periodic
function on the atomic scale.  This identity can be proved
by assuming that $f({\bf r})$ is a constant in each unit cell
of the crystal.  The set of equations for the envelope functions
is then (assuming an AlGaAs-GaAs-AlGaAs type
heterostructure in the z direction, with A/B below denoting 
AlGaAs-GaAs)
\begin{eqnarray}
\sum_m \left\{ \left[ \frac{{\bf p}^2}{2m_0} + \left(
\epsilon_l^A Y_1 + \epsilon_l^B Y_2 + \epsilon_l^A Y_3 
+ V({\bf r}_{\perp}) \right) \right] \delta_{lm} \right. & & 
\nonumber \\
\left. + \left( {\bf \pi} \cdot {\bf p} \right)_{lm}
\right\} f_m ({\bf r}_{\perp}, z) = E f_l ({\bf r}_{\perp}, z) 
& & 
\,.
\end{eqnarray}
Here $\epsilon_l^A$ and $\epsilon_l^B$ are band edge 
energies of materials A and B at the $l$th band edge 
at $\Gamma$ point.  $Y_i$ are step functions that take 
value 1 for the $i$th layer
and 0 everywhere else---we assume sharp interfaces
between materials A and B (although deviations from
sharpness can be built into the model).  
${\bf \pi}$ is the interband 
transition matrix, which is essentially the expectation
value of momentum operator ${\bf p}$ in a unit cell.  
We can separate f into an in-plane 2D and 
the z direction components and further simplify the
equations.  In addition, the valence band 
envelope functions can be written in terms of the two 
conduction band functions, thus leading to a {\it nonlinear}
(but only 2 by 2) eigenvalue problem.
The presence of a slowly varying electric
field (for the purpose of confinement) plus the band edges 
for the heterostructure introduces additional (on top of
interband coupling in bulk GaAs)
coupling between different bands \cite{Bastard}.  At the 
zeroth-order approximation, 
when one neglects all spin-orbit 
coupling terms and interband couplings, the set of 
equations above reduces to the single envelope
function Schr\"{o}dinger equation we employ in our current study.
In our approximation the only effects of the band structure 
are to replace the bare electron mass $m_0$ by an effective 
mass $m^*$ and the bare Coulomb interaction by its screened
form, which is precisely the single envelope function 
effective mass approximation.  

To validate our zeroth-order approximation, 
we need to estimate the magnitudes of the higher order 
corrections neglected in our approximation.
In particular, we can evaluate 
the following quantity $p=E_p \bar{E} m^* / m_0 E_g^2$ 
\cite{Bastard}, where $p$ is 
the strength of the interband and spin-orbit corrections 
relative to the zeroth-order terms within each conduction band.
Here $E_p$ represents the interband coupling 
strength, $\bar{E}$ is the characteristic
electron envelope energy, $m_0$ is the bare electron mass, 
$m^*$ is the conduction band effective mass, 
and $E_g$ is the fundamental band gap at the $\Gamma$ point.    
For GaAs, $E_p = 
22.71$ eV; $m^* = 0.067 m_0$, $E_g = 1.5192$ eV \cite{Bastard}, 
and the characteristic electron energy $\bar{E}$ 
is about $10$ meV.  Using these parameters, we 
obtain $p \sim 1/150$, which is indeed a
small quantity, justifying our envelope function effective
mass approximation in the low energy singlet/triplet sector.  
The off-diagonal corrections, which couple
the spin up and down components of the envelope wavefunctions,
have similar negligibly small magnitudes.  
For the spin-coupling the corresponding small
parameter is $p^{\prime} = E_p \Delta \bar{V} m^* 
/ m_0 E_g^3$, where $\Delta$ is the valence band splitting 
due to spin-orbit coupling and $\bar{V}$ is the average 
confinement energy.  For GaAs $\Delta = 0.341eV$, while we
take $\bar{V} \sim 50meV$.  We then obtain $p^{\prime} \sim
p \sim 1/150$.
Therefore, up to an accuracy of 1\%, the conduction bands 
of two different spins are decoupled from each other and from 
other valence bands, and the use 
of a single band envelope function may be quite useful
qualitatively and semi-quantitatively.  It would, however,
require further numerical investigations going beyond the
single envelope function approximation to establish
whether this accuracy is consistent with the stringent 
error correction requirements in QC.

To go beyond the zeroth-order approximation, the above-mentioned
correction terms need to be included, and the linear 
Schr\"{o}dinger equation we have now becomes a nonlinear
eigenvalue problem, with a non-vanishing off-diagonal 
term that couples the up and down spins.  Thus, strictly
speaking, because of spin-orbit coupling,
the spin up and down states of a conduction electron
are not the eigenstates in a semiconductor heterostructure.
This opens another possible, albeit weak, channel for spin 
decoherence in quantum dots that is not present in the bulk.

When a magnetic field is introduced, it can be directly
incorporated in the envelope function effective Hamiltonian.
The underlying $\Gamma$-point Bloch functions, which are
atomic wavefunctions, are only minimally affected by the
external magnetic field.  Indeed, in a typical atom, the
${\bf A}^2$ term is about $10^{-3}$ as big as the linear
term in a 10 Tesla field, which can be safely neglected.  
The linear term
in ${\bf A}$ corresponds to the coupling between the electron
orbital angular momentum with the external magnetic field.
For the S orbital of the conduction band, this coupling
vanishes; for the P orbitals of the valence bands, 
the magnitude of the splitting caused by this term is 
about 1 meV per 18 Tesla. Compared to the main gap of 
about 1.5 eV, this splitting can also be safely dropped.
Therefore, we can conclude that the underlying Bloch
functions are not affected by any moderate (up to 10 Tesla)
external magnetic fields one needs for QDQC operation.
We also conclude that for the purpose of QDQC operations, 
where one restricts to only the low energy singlet/triplet
sector of the Hilbert space, the single envelope function 
effective mass approximation employed in this paper is 
qualitatively excellent, but further studies are needed to 
establish whether this approximation satisfies the demanding 
constraints of error correction in a realistic QDQC 
architecture.

\subsection{Singlet-triplet crossing}

In our calculations we find a singlet-triplet crossing in all
the situations we considered for a magnetic field around
4 Tesla.  The physical reason underlying this magnetic
field induced singlet-triplet crossing (making the triplet
state the ground state in high fields) is somewhat subtle.
In a single quantum dot ``helium atom'' (two electrons in one dot), 
where such a crossing has also been reported in the
literature, the competition between inter-electron
Coulomb repulsion and single particle 
excitation is the reason for the singlet-triplet
crossing \cite{Ashoori,Wagner}.  In a quantum dot 
hydrogen molecule that we consider (two electrons in 
two dots), the
electrons can reside in different dots, minimizing
the Coulomb repulsion effect, and therefore the above
reasoning does not really apply for our singlet-triplet
crossing.  To achieve a better
understanding of this crossing, we first write down the
expression for the singlet-triplet energy splitting (exchange
coupling) using the Heitler-London wavefunctions of 
the ground singlet and triplet states: 
\begin{eqnarray}
|\Psi_s \rangle & = & \left[ \frac{|\phi_L(1)\rangle 
|\phi_R(2)\rangle
+ |\phi_L(2)\rangle |\phi_R(1)\rangle}{\sqrt{2}} \right]
\nonumber \\
& & \times \frac{|\!\uparrow \downarrow \,\rangle 
- |\!\downarrow \uparrow \,\rangle}{\sqrt{2}} \,, \\
|\Psi_t \rangle & = & \left[ \frac{|\phi_L(1)\rangle 
|\phi_R(2)\rangle
- |\phi_L(2)\rangle |\phi_R(1)\rangle}{\sqrt{2}} \right]
\nonumber \\
& & \times \frac{|\!\uparrow \downarrow \,\rangle 
+ |\!\downarrow \uparrow \,\rangle}{\sqrt{2}} \,,  
\end{eqnarray}
where $|\phi_L\rangle$ and $|\phi_R\rangle$ are localized electron
spatial orbitals.  The exchange coupling---the energy splitting 
between the lowest triplet and singlet states---can then be 
expressed as
\begin{eqnarray}
J & = & \frac{\langle \Psi_t|H|\Psi_t \rangle}
{\langle \Psi_t|\Psi_t \rangle} 
- \frac{\langle \Psi_s|H|\Psi_s \rangle}
{\langle \Psi_s|\Psi_s \rangle} \nonumber \\
& = & J_r + J_c \,, 
\end{eqnarray}
where $J_r$ is the contribution from the single particle potential 
energy, while $J_c$ is the contribution from Coulomb interaction 
between the two electrons.  $J_r$ and $J_c$ can be expressed as
\begin{eqnarray}
J_r & = & \frac{2|S_{LR}|^2}{1-|S_{LR}|^4} \left[ \langle \phi_L|
\Delta V_L |\phi_L \rangle + \langle \phi_R|
\Delta V_R |\phi_R \rangle \right. \nonumber \\
& & \left. - \langle \phi_R|
\Delta V_L |\phi_L \rangle - \langle \phi_L|
\Delta V_R |\phi_R \rangle \right] \,, \\
J_c & = & \frac{2|S_{LR}|^2}{1-|S_{LR}|^4} \left[ \langle 
\phi_L(1)\phi_R(2)|e^2/\epsilon r_{12}|\phi_L(1) 
\phi_R(2)\rangle \right. \nonumber \\
& & \left. - \frac{{\rm Re}\langle 
\phi_L(1)\phi_R(2)|e^2/\epsilon r_{12}|\phi_L(1) 
\phi_R(2)\rangle}{|S_{LR}|^2}
\right] \,, 
\end{eqnarray}
where $S_{LR} = \langle \phi_L| \phi_R \rangle$,
$\Delta V_L = V(x,y) - V_L$,
and $\Delta V_R = V(x,y) - V_R$, with 
$V(x,y) \equiv V({\bf r}_i)$ of Eq.~(\ref{eq:Hamiltonian}).
Here $V_L$ is a harmonic well located on the left and $V_R$ is a 
harmonic well located on the right.  The basis wavefunctions
$\phi_L$ and $\phi_R$ are eigenstates of these two wells
respectively.  Thus, we can see that $J_r$ is a contribution 
due to the difference caused by replacing the actual 
confinement potential $V$ by a left or right harmonic well.
It is a single particle contribution.
Whether $J_r$ is positive or negative depends on the particular
choice of $V$ and the parabolicity choice for $V_L$ and $V_R$.
When the distance between the two quantum dots becomes large, 
this quantity approaches zero.

The Coulomb contribution $J_c$ consists of two parts,
one from direct Coulomb interaction, the other from 
exchange interaction.  These two parts generally do not have the
same type of dependence on an external magnetic field ${\bf
B}$.  As ${\bf B}$ increases in strength, the exchange
interaction becomes more important, which leads to the
singlet-triplet crossing in a quantum dot molecule.  An
analytical calculation for a special (somewhat artificial) 
confinement potential has been recently performed \cite{BLD},
which explicitly demonstrated the different behavior of
the direct Coulomb and the exchange terms.

The expression for $J$ shows that there are multiple contributions
to the energy difference between singlet and triplet states.
Without analytical expressions it is difficult to determine
exactly which factor dominates in a particular range of parameters.
Physically, the Pauli principle constraint determines
that in a triplet state the two electrons will try to avoid
each other, thus establishing a repulsive correlation between
them.  This correlation helps to lower the Coulomb interaction 
energy, favoring the triplet state to have
lower energy if Coulomb interaction is dominant.  
As the external magnetic field is increased,
the wavefunction overlap decreases because of the squeezing
by the magnetic field,
so the long range Coulomb interaction becomes the dominant
factor in the total two-electron interaction energy, leading to
the triplet state being the ground state at high enough
magnetic fields.  On the other hand, at lower magnetic fields,
the wavefunction overlap is significant, a singlet
state is then the ground state since it lowers the
electron kinetic energy.
A singlet-triplet crossing is therefore inevitable as
a function of the magnetic field, which for the double-dot
parameters we choose, happens at rather low accessible 
fields of 4 Tesla.

\subsection{Quantum chemical approaches}

As we mentioned before, the Schr\"{o}dinger equation for a 
two-center two-electron problem cannot be solved exactly.  
Various quantum chemical approximations have been proposed and
implemented in this problem in the context of electronic energy 
level calculations in real molecules \cite{McWeeny}.  
Below we present a summary and a critique of the various
techniques which may be useful in the calculations for 
obtaining realistic QDQC architectural parameters.  
Since the exact electron wavefunctions
are important in the context of quantum computing, we will
not discuss approaches that deal only with electron 
charge or spin densities.  We believe that detailed 
electronic structure calculations, which provide accurate
information about the wavefunctions spanning the relevant
Hilbert space for realistic QDQC architecture, will be 
absolutely essential for further progress in this field.

\subsubsection{Hartree-Fock approximation}

One of the simplest quantum chemical approaches is the
Hartree-Fock (HF) approximation.  
It uses an effective single-electron equation 
to simulate a two-electron problem. Pauli principle is accounted 
for because the two-electron wavefunction is written as a 
single Slater determinant.  
The HF equation can be solved directly or on a finite basis
(the so-called linear combination of atomic orbitals 
method, abbreviated as LCAO).  The main advantages of a 
HF calculation are its single particle feature, its
accessibility, and its clear underlying physical picture.  
The main shortcoming is its disregard of electron correlations, 
which originates from the
simplification of a two-particle problem to a one-particle 
problem.  This deficiency can be systematically
remedied by introducing configuration interaction (CI)
corrections.  Instead of using a single Slater
determinant as the system wavefunction, one can use a series of 
Slater determinants (in which the single particle wavefunctions 
are HF wavefunctions including the excited states) as 
basis and search for the best combination.  As the size of this 
basis goes to infinity, the method becomes exact.  One may, 
however, be able to obtain very high accuracy with a 
reasonable size CI calculation if the configurations to be 
mixed in are chosen judiciously.

One potential shortcoming of the Hartree-Fock method for the 
purpose of quantum computation is that it may not be
sufficient to describe quantum entanglement.  
Multi-electron wavefunctions are intrinsically inseparable
when there are overlaps between single electron 
wavefunctions.  As a consequence
of electrons being {\it indistinguishable}, a Slater
determinant is not a simple separable product function, 
and therefore individual electron states generally 
cannot be factored.  For example, consider a two
particle Slater determinant:
\begin{equation}
|\Psi\rangle = \frac{1}{\sqrt{2}} \left[ 
|\phi(1)\rangle |\!\uparrow\rangle_1
|\psi(2)\rangle |\!\downarrow\rangle_2 - 
|\phi(2)\rangle |\!\uparrow\rangle_2
|\psi(1)\rangle |\!\downarrow\rangle_1 \right] \,. 
\label{eq:SlaterDet}
\end{equation}
We can easily calculate the single particle density matrix for
particle 1:
\begin{eqnarray}
\rho_1 & = & {\rm Tr}_2 (\rho_{12}) = {\rm Tr}_2 
(|\Psi\rangle \langle \Psi|) \nonumber \\
& = & \frac{1}{2} \left[ |\phi(1)\rangle \langle\phi(1)|
|\!\uparrow\rangle_1 \langle\uparrow \!|_1 +
|\psi(1)\rangle \langle\psi(1)|
|\!\downarrow\rangle_1 \langle\downarrow \!|_1 \right] \,,
\end{eqnarray}
which is indeed a mixed state.
However, this inseparability in the Slater determinant 
arises only from
correlations due to the Pauli exclusion principle.
If the electrons are spatially separated so that they
become {\it distinguishable}, the 
electron wavefunction of Eq.~(\ref{eq:SlaterDet})
becomes a product.  For example, if $\phi$ and $\psi$ 
above are localized spatially with no overlap,
the above two-particle wavefunction simplifies to
\begin{equation}
|\Psi\rangle = 
|\phi(1)\rangle |\uparrow\rangle_1
|\psi(2)\rangle |\downarrow\rangle_2 \,. 
\label{eq:purestate}
\end{equation}
Here 1 and 2 are labels of the two 
distinguishable particles---particle 1 in
$\phi$ and particle 2 in $\psi$.  The 
two-electron wavefunction is 
now in a product form and the state for each particle
is pure.  In other words, Eq.~(\ref{eq:purestate}) is
not an entangled state.

In the RHF approach, 
the spin part of the wavefunction would be a singlet for two 
electrons when they occupy the same spatial orbital, so that
the state is necessarily
entangled.  The entanglement here is fundamentally different
from the inseparability 
that arises purely out of the Pauli exclusion principle 
as considered above.  Instead, it represents 
a true correlation between the two particles---they occupy
the same spatial orbital.
On the other hand, in the UHF approach the wavefunction
is completely separable when the two wavefunctions are localized, 
so that no entanglement can be described.
In general, for indistinguishable particles, 
the entanglement information is encoded in the form
of superposition of different Slater determinants so that,
in principle, CI is always {\it needed} for the 
wavefunction to carry entanglement information.
From another perspective, for $n$ spin-$\frac{1}{2}$ particles,
the number of real variables needed 
to describe the spin part of the entangled multi-electron
wavefunction is $2^{n+1}-2$, while one Slater determinant 
only provides $2n$ real variables to describe the spin degrees
of freedom, which is clearly not enough to incorporate 
entanglement in any multi-electron case, including even
the n=2 two-electron case we consider here.
Therefore, the single Slater determinant HF approximation
is manifestly inadequate for QC purposes.
One should note, however, that the HF approximation is
not intended for the purpose of describing quantum entanglement.
It is designed to compute accurately electronic energy spectra,
charge and spin densities, etc.  Therefore, as long as one
recognizes the shortcomings of this method, it can still
provide valuable information about the electronic system.

\subsubsection{Heitler-London
method}

Hartree-Fock calculation is self-consistent, in which the mean
field is produced by the calculated electron density.  One can
also solve the two-electron problem using a fixed finite 
molecular orbital basis.  Indeed, when the 
number of states in the basis goes to infinity, the 
solution approaches the exact two-electron state.  However, the
convergence may be slower than a self-consistent 
calculation (with CI), and it quickly becomes computationally 
intractable for multi-electron problems.  On the other hand, for a 
two-electron problem with a small number of basis states,
such a fixed finite basis calculation is numerically 
tractable and provides a clear advantage over the HF 
approximation for studying entanglement.

Heitler-London method is an approximation to
the simplest molecular orbital calculation.
Here only the two single particle ground states in the individual
quantum dots are taken into account.  Furthermore, in forming 
two-electron orbitals, the two ``polar'' (doubly occupied) states
are neglected.  There are then only one possible 
functional form(s) for
singlet and triplet states respectively.  This approach is quite
accurate when the two dots are far from each other, so that the
single particle wavefunctions have the correct dependence on
the inter-dot distance.  
On the other hand, if the two dots are brought close to 
each other, the wavefunctions' radii should be varied in order to
obtain the lowest energy for the two-electron states.  This is
similar to the orbital contraction in molecular physics when two
binding atoms are brought together \cite{McWeeny}, although in
quantum dots it might be orbital expansion
rather than contraction.  Another way to 
improve Heitler-London calculation is to introduce orbital
``polarization'' (contortion of the s-orbitals towards each
other) so that anisotropies in the problem can be at least 
partly accounted for.  For example, p-orbitals can be included
in the single particle wavefunctions (s-p hybridization) 
\cite{BLD}.  Indeed,
in the case we considered in this manuscript, s-p hybridization
is an extremely important feature of the problem, as we already 
discussed in sections \ref{sect:theory} and 
\ref{sect:results} of the paper.

\subsubsection{Molecular orbital theory}

Heitler-London method is appealing in its simplicity and its
clear physical picture.  However, unless perfect basis states 
happen to be ``luckily'' chosen, it is difficult for a method
with such a small basis to accurately describe a double-dot
molecular system.  The first improvement one can make is to
include the polar states.  It then becomes the simplest molecular 
orbital calculation---the Hund-Mulliken approach \cite{BLD}.  To
further enlarge the basis, one just includes more single particle
orbitals.  For example, in our case, we have included all the single 
particle first excited states, so that there are in total 6 single
particle states forming our basis, 
from which we can form 36 two-particle states.
If $n$ single particle orbitals are used, the number of two-particle
states grows as $n^2$, while the number of Coulomb matrix elements
grows as $n^4$.  It is thus imperative
to select the best possible single particle wavefunctions so that
the number of these orbitals can be kept a minimum, allowing
a tractable computation.

To limit the size of the two-electron state basis, one can select
a portion of states from a more complete basis, using criteria 
such as single particle energy cut-off.  Such an approach
amounts to a Heitler-London calculation supplemented by limited
CI.  However, there can always be hidden
hazards in this practice.  For example,
as has been pointed out \cite{Herring}, the
calculation of exchange energy is nontrivial in an array of
atoms.  One reason is that exchange is mainly determined
by tail overlaps between neighboring electron wavefunctions,
where Heitler-London wavefunctions (often used for calculating
exchange energy) is less reliable.  In addition, including more
configurations and going beyond Heitler-London wavefunctions may
not improve the accuracy because the excited atomic
wavefunctions have different exponential tails.  Thus the
eigenstates may have more accurate shapes near the atomic
cores, but their tails may become less accurate, leading to
inaccurate exchange coupling energy.  Indeed, it is
always a dangerous practice to obtain a small quantity 
numerically from the
difference of two large quantities.  In quantum dot molecules,
the tail behavior of wavefunctions is somewhat simpler than in
atoms because all the harmonic well eigenfunctions have the same
long distance exponential behavior 
(but multiplied by different polynomials).  Therefore, by
including a larger basis and doing limited CI
calculations, one should be able to achieve a reasonable
description of the eigenstates, eigenenergies, and in
particular, the exchange energy in QDQC architectures.
In this particular sense, the 2D harmonic confinement in
QD systems may provide a significant calculational advantage
over the corresponding real atom/molecule situations
with Coulomb confinement.  On the other hand, the non-singular
nature of the harmonic potential well also means that the
electronic states are more sensitive to the actual details of 
the confinement, making QDQC architecture a fragile one
for quantum computation.

\subsubsection{Hubbard model}

The Hubbard model is a highly simplified model describing
Coulomb correlation effects in an array of atoms.
The model \cite{Auerbach} deals with a second 
quantized multi-electron 
Hamiltonian with a cut-off in the interaction.  
Only one orbital(s) state per site is kept (in the atomic 
limit), and there is a finite hopping term $t$, 
arising from overlap, between the neighboring orbitals. 
The long range Coulomb interaction is replaced 
by a single on-site repulsion term $U$---the rationale
being that screening by all the other electrons
lead to an on-site effective $U$.  The 
ferromagnetic direct exchange part of the Hamiltonian is 
dropped because the wavefunction overlap between neighbors 
is exponentially small.  Multi-site Coulomb interaction
is also neglected, assuming that they do not affect the magnetic
properties of the model.  In the limit of large on-site
repulsion (large $U$), the effective Hamiltonian that 
describes the excitation of this model is a Heisenberg 
exchange Hamiltonian, with the exchange coupling $J$ 
related to $t$ and $U$ by $J=4t^2/U$.  As this $J$ is
always positive, the ground state is antiferromagnetic
when there is one electron per site.  
There are various attempts to add additional terms to
the Hubbard model (the extended Hubbard models)
so that it can describe various other
phenomena.  For example, in one extended 
Hubbard model, nearest neighbor Coulomb interaction is also 
taken into consideration.  The model can then describe 
spatial charge density fluctuations. 

The Hubbard model and its variants have been applied
to quantum dot arrays \cite{Kotlyar},
particularly in the context of transport and magnetic 
properties and also to study quantum phase transitions 
in quantum dot arrays.  It is an effective model that 
can describe complex phenomena with simplicity.
In the context of QDQC using spins as qubits it is
unclear that Hubbard model could have much relevance
because of its extreme simplicity.  This is certainly
true for the two-electron in the double-dot problem 
studied in this paper.
However, if multiple-dot algorithms are designed
in the future,
the Hubbard model may become a powerful tool, although various
details will have to be added,
diminishing the simplicity of the original model.

The Hubbard model reduces to the Heisenberg model in 
the large on-site repulsion limit ($U \rightarrow \infty$).
One condition for the validity of the
Heisenberg exchange Hamiltonian is that each localized electron
wavefunction should have exponentially small overlaps with
others.  This condition is generally not satisfied 
in QDQC when we bring
two quantum dots very close to each other.  However, for a 
two-electron problem, if the orbital degrees of freedom are frozen,
the spin degree of freedom has only four dimensions which can be
described by the singlet and triplet states, and a Heisenberg
model description becomes possible.  On the other hand,
if the electron orbital degrees of freedom are ever excited, the
Heisenberg exchange Hamiltonian will not be applicable for our
purpose. For example, if two electrons ever get into one quantum
dot simultaneously, 
we will lose track of which qubit is represented by which
spin, thus error probability would be 50\%.  As we have shown
previously, at low magnetic fields in the current configuration,
the ground singlet state has a finite probability (as large as
20\% or more at zero magnetic field) of double occupation in
either dots.  For a short QC calculation or for qualitative 
results, a 5\% double occupation probability may be acceptable.  
However, this small error becomes a very serious
problem that cannot be overcome by currently available
error correction schemes for a long quantum computation,
leading to the constraint that the double occupation 
probability must be kept very low.

\subsection{Various aspects of a quantum dot quantum computer}

It has been pointed out that the spin-based quantum dot 
quantum computer, in principle, satisfies the necessary 
and sufficient conditions required for quantum computing \cite{LD}.  
Here we would like to discuss in further details
a number of salient features that arise naturally in
the context of QDQC.  

In many materials, electron spins are less vulnerable to 
decoherence than their orbital degrees of freedom,
which in fact is the main motivation for the proposed
spin-based QDQC.
For example, carriers in GaAs have a sub-picosecond momentum 
relaxation time while their spin relaxation time is longer
than one nanosecond \cite{Awschalom}.  Since long spin
coherence time is absolutely essential for QDQC operations
(in particular the spin coherence time must be much longer 
than the single- and two-qubit operation times, which have 
to be controlled by switching magnetic fields and gates and
cannot be very fast) we briefly
summarize spin relaxation mechanisms and comment on their 
relevance in the context of 2-dimensional GaAs quantum dot
structures.

There are three major spin relaxation channels for conduction 
electrons in GaAs: 
the Elliott-Yafet mechanism (EY), the D'yakonov-Perel' 
mechanism (DP), and the Bir-Aronov-Pikus mechanism (BAP) 
\cite{Fabian2}.  The EY process originates from the fact that 
real Bloch functions are not spin eigenstates.  For example, 
spin-orbit coupling can mix spin up and down states in the 
electron eigenstates.  Thus, whenever an electron is scattered
(by other electrons, phonons, impurities, etc.), 
there is a finite probability that the dominant spin component 
will flip, thus causing spin relaxation.
The DP channel arises from the lack of inversion symmetry
in GaAs, which leads to an intrinsic spin splitting in
the conduction band even for zero magnetic field.
In the DP channel, the energy band splitting due to spin-orbit  
coupling is treated as an effective magnetic field.  
For different ${\bf k}$ states, this effective field has
different magnitudes and directions.  Thus, as an electron is 
scattered from one momentum state to another, it sees different 
fields and precesses differently whenever it is scattered.  
Soon the electron loses the memory of its initial spin state, 
thus showing spin relaxation.
The BAP channel is somewhat similar to the DP channel, as it also
treats electron spins as precessing in an effective magnetic field.
However, in the BAP mechanism the effective field 
for the conduction electrons is produced by
free or bound holes (or other paramagnetic impurities which
may be present).  Hole spins relax very 
fast, so that the effective fields (the the conduction electrons)
produced by hole spins fluctuate,
which causes an electron spin to lose the information of its 
initial state.

In GaAs heterostructures, it is generally believed that the
DP mechanism is the dominant spin relaxation channel
for conduction electrons \cite{Fabian2}.  In particular,
due to the band discontinuity in a heterostructure, 
there is an additional
spin splitting for the conduction electrons (``Rashba'' effect)
which can be treated as an extended DP channel.
For holes, however, the EY mechanism is the dominant process.
An additional complication is that 
in a quantum dot produced by modulating electric fields through
lithographic gates, 
the confining electric field produces a mixing 
between the spin up and down states (in addition to the basic 
splitting arising from the lack of inversion symmetry in GaAs).  
The boundaries and the interfaces are also known to
cause spin relaxation.  Indeed, 
these additional spin relaxation channels
may actually be the dominant processes for the
electrons confined in the quantum dots, because the
wavefunctions for these electrons are built from
the $\Gamma$ point Bloch functions, 
where the underlying Bloch function is 
S type which has no spin-orbit coupling.
Since the DP channel
depends on the effective field produced by the spin-orbit
coupling (which vanishes at the $\Gamma$ point), 
and an external magnetic field may also 
help limit the DP channel, 
electron spin relaxation in a quantum dot
should be quite weak (and probably arises primarily from
the interface/boundary scattering, the
confining electric field, and perhaps the Rashba effect).  

When electron spin relaxation originating from the spin-orbit 
coupling (DP channel) is largely suppressed, other 
relaxation channels 
have to be carefully considered.
In particular, interface/boundary scattering induced 
spin relaxation needs to be considered.
In addition, it has been pointed out \cite{BLD} that
the nuclear spins may affect electronic spin relaxation
through the hyperfine 
interaction.  This spin relaxation channel can, however,
be substantially suppressed
by applying an external magnetic field or through the
Overhauser effect to dynamically polarize the nuclear spins
\cite{BLD}.  These issues require more careful (and
quantitative) considerations before QDQC architecture 
questions can be seriously considered.

If the spin-orbit coupling is strong so that electron 
spin by itself is no longer a good quantum number,
then one must consider the total angular momentum J, 
which involves 
both spin and orbital degrees of freedom.  Such a mixing 
by itself would not be a disaster for quantum computing
since J can now replace electron spin to serve as 
the qubit.  However, the ``spin'' relaxation
time will then be in the same order of magnitude
as the momentum relaxation
time, which is generally very short ($\sim$ ps or less) 
in semiconductors, which could be disastrous from the 
QC perspective.  
It is thus imperative to choose materials
with small spin-orbit coupling for the purpose of electron spin
quantum computing---otherwise decoherence problem makes
QC operations impossible.

Even if spin relaxation can be neglected (because of,
for example, long spin coherence time), there are 
many other factors that can affect the performance of a
quantum computer.  For example,
based on our molecular orbital calculation, the exchange 
coupling in a coupled dot system is found to be large 
enough for fast 
picosecond switching.  However, one has to be careful 
in exploiting the possibility of fast switching.
Indeed, to produce the best structure for the purpose of 
quantum computing, a compromise needs to be achieved between 
an optimal gating time and an optimal 
error rate that should both be as small 
as possible.  As we learned from our calculations and from
general arguments, exchange coupling decreases exponentially 
fast as the two dots are pulled apart.  Consequently, to have 
a larger exchange coupling (which means a shorter gating time),
the dots should preferably be close to each other, 
which, however, increases the
overlap between the electron wavefunctions,
leading to enhanced double occupation probability, which
means higher error rate.  A compromise in the QDQC architecture 
will therefore be needed.

As shown in our molecular orbital calculation,
the subspace of the total Hilbert space containing
the ground singlet and triplet states is well separated from 
the rest of the Hilbert space, and can thus be treated as an
isolated system.  This is the whole idea behind using the
exchange coupling for the purpose of quantum gating.  Moreover,
as long as the Heisenberg Hamiltonian can be used to describe
the quantum dot two-spin system, the spin singlet and 
triplet states are always the exact eigenstates.  
The only important parameter for state
evolution is the time integral of the Hamiltonian 
$\int H(\tau) d\tau$, and the exchange coupling should be turned on 
for as short a time as possible to produce an ultrafast gate.  
However, this time cannot be too short as to make the system 
``leaky''.  Using the uncertainty principle, we can estimate the
lower limit of this turn-on time $\tau_p$.  Recall that the next
excited state of our two-electron system is about 8 meV above 
the ground states.  Thus, the lower limit of $\tau_p$ is about
($\delta E \sim 8$ meV is the energy difference between the 
next excited state and the ground singlet and triplet states)
\begin{equation}
\tau_p \gg \hbar/\delta E \sim 0.1 \ {\rm ps} \,.
\end{equation} 
Therefore, as long as the gating time $\tau_p$ 
is longer than 1 ps
in the current configuration, the coupled dot system is 
well isolated, so that the higher excited states can be 
safely neglected, and the gating action can be considered 
adiabatic.  This is critical for QC operation.  Again
a compromise is needed here to optimize a fast gating
time and adiabaticity.  Calculations of the kind carried 
out in our paper can provide quantitative estimates for 
such required QDQC architectural optimization.

According to Fig. 6, the exchange coupling $J$ depends quite
sensitively on the magnetic field ${\bf B}$.
If a sequential pulse algorithm is used, one does not
need to worry about the interplay between the exchange 
interaction and the local magnetic field.  On the other hand, 
if a parallel pulse scheme is used \cite{parallel}, one has 
to take into consideration the effect of the inhomogeneous
magnetic field on the exchange coupling.
Intuitively, the average field exchange coupling may be
sufficient in many cases, because the single electron
wavefunction radius decreases slowly as the magnetic field 
increases: $l_B = \sqrt{\hbar c/e B}$.  If the average field
is around zero, the field inhomogeneity may lead to a bigger
change in the exchange coupling, and will have to be taken
into account.   

Throughout our calculation we have neglected the 
Zeeman splitting of the electronic levels.  This splitting
cannot be ignored in a real unitary evolution.  For instance,
in Ref.~\cite{BLD} a pulse sequence was given for a 
controlled-NOT (CNOT) gate (the sequence as given is a 
conditional phase shift, which can be easily transformed 
into a CNOT operation).  If a finite ${\bf B}$ field is 
present during the
swap action, an additional phase due to Zeeman splitting
of the triplet states will show up in the electron spin
states.  An opposite ${\bf B}$ field with the same strength has to 
be applied to the two electrons for the same amount of time
as the swap gate to correct this phase error.  For GaAs, the 
Zeeman splitting is about 0.03 meV/Tesla.  If a dot system 
has an exchange coupling of 0.1 meV and the two spins experience
a magnetic field difference of 1 Tesla, the corresponding 
difference in the Zeeman splitting would be 0.03 meV, 
about 30\% of the exchange coupling, which is a significant
number.  
Since an error rate below $10^{-4}$ is needed for the currently 
available error correction schemes to be effective, one has to 
be able to control the magnitude of the exchange interaction 
up to that accuracy.  Furthermore, any residual local field effect 
has to be corrected continuously.  Indeed, if in an actual 
structure the gating area is separated from the storage area, 
which means that all the spins have to be transported to and 
from the gating area, one does not need to worry about the stray
magnetic field.  Here the main problem would be the 
transportation of the spins.  On the other hand, if the spins 
are stored close to each other and the gating and storage 
areas are combined, the main problem would be the effects of 
the local stray magnetic field.  It is straightforward to correct 
for the effect of a magnetic field on one spin.  However, it is 
much less obvious how to correct for the effect of an inhomogeneous 
field on all but one spin in a chain.  From an engineering 
perspective, the modular approach of separating storage and 
gating areas is somewhat more promising.  We anticipate that
the inhomogeneous field and the stray field problems to be
significant obstacles in producing a successful QDQC 
architecture.

When electron transport is needed in an architecture, electron 
labeling becomes very important.  In a semiconductor heterostructure,
there always exist stray electrons, such as those trapped in
impurities and deep levels.  
If we move our qubit electron around in a heterostructure,
there is the danger of losing this electron, 
and in its place, acquire a
stray electron, so that all the spin information of the 
particular qubit is lost.  The indistinguishable character of
electrons becomes an important issue in this context.  Initially,
when all the electrons are trapped in their respective quantum dots,
they are labeled and distinguishable.  As soon as stray electrons
are present outside the dot electrons we have considered, 
Pauli exchange errors will occur from the indistinguishability
of fermions and have to be corrected \cite{Ruskai}.  This will
be another significant obstacle for the QDQC architecture.

Experimentally, it is easier to deal with multiple electrons 
(instead of a single electron) in a 
quantum dot produced by modulating electric fields.  Here it is
hoped that certain shell structures exist 
(as we show in our results) so that such a quantum
dot can be considered to be an effective spin-$\frac{1}{2}$ system.
Multiple electrons may, however,
make the indistinguishability problem a more
prominent issue.  However, one needs to keep in mind that the
important question here is the spin state of the effective 
spin-$\frac{1}{2}$ system, not the spin state of any particular
electron.  We are currently pursuing multi-electron 
calculations in order to better understand these difficult and 
complex issues. 

If the exchange coupling J is tuned by changing 
external gate voltage in a QDQC, thermal fluctuations 
(or any other types of fluctuations) in the gate voltage will
lead to fluctuations in J, thus causing phase errors in the
swap gate that is crucial for two-qubit operations.  Here we
estimate this error by assuming a simple thermal (white) 
noise \cite{Bkane1}.

Assuming that $J=f(V)$ where $J$ is the exchange coupling and
$V$ is the gate voltage that controls the value of $J$, around
any particular value $V_0$, $J$ can be expressed as $J(V) =
J(V_0) + \left. f^{\prime}(V) \right|_{V_0} (V-V_0)$.  During a
swap gate between two quantum dots, the phase of the electronic 
spin wavefunction evolves as $\phi = \int_0^t J(\tau) d\tau /\hbar$.
Thus the fluctuation in the phase $\phi$ is 
\begin{eqnarray}
\langle \delta \phi^2 \rangle & = & \langle \phi^2 \rangle
- \langle \phi \rangle^2 \nonumber \\
& = & \frac{1}{\hbar^2} \int_0^t \int_0^t \langle \delta 
J(\tau_1) \ \delta J(\tau_2) \rangle d\tau_1 d\tau_2 
\nonumber \\
& \sim & \int_0^t \int_0^t 
\frac{[f^{\prime}(\bar{V})]^2}{\hbar^2}
\langle \delta V(\tau_1) \ \delta V(\tau_2) \rangle 
d\tau_1 d\tau_2 \,.
\end{eqnarray}
If $|f^{\prime}(\bar{V})|$ is bounded by a constant $\alpha$
we can replace it by $\alpha$ in the above expression.  
Furthermore, according to Nyquist theorem, 
\begin{equation}
\langle \delta V(\tau_1) \ \delta V(\tau_2) \rangle
= 4 R k_B T \delta(\tau_1-\tau_2) \,.
\end{equation}
Here $R$ is the circuit resistance and $T$ is the circuit 
temperature.
We thus obtain the approximate expression for the phase 
fluctuation:
\begin{equation}
\langle \delta \phi^2 \rangle
\sim 4 R k_B T \alpha^2 t /\hbar^2 \,.
\end{equation}
In our calculation for double quantum dot QC architecture, 
$V_b$ plays the role of external gate
voltage.  According to Fig. 9, in the two higher barrier 
cases, $J$ changes about 0.038 meV when $V_b$ 
(the strength of the barrier Gaussian, not the effective
barrier height) changes
1.83 meV.  $\alpha$ can be obtained from this ratio
as 0.021 eV/V.
Assuming the swap gate is performed at 1 K
(since J is in the order of 0.1 meV $\sim$ 1 K, the
experimental temperature can only be lower than 1 K),
and the transmission line connecting the gate to the outside
has an impedance of 50 ohm, the rate for phase fluctuation
$\langle \delta \phi^2 \rangle/t$ is about 3.2 MHz.  The 
phase error accrued during a swap gate is about 
0.06\%.  This is quite small an error which is in the same
order of magnitude as the theoretical tolerance of the 
currently available quantum error correction codes.
It should pose no problem to any demonstrative experiment.  
For a real quantum computer, 
this error rate needs to be further lowered by lowering 
experimental temperature 
and turning up $J$ more gently (which requires longer time 
but produces smaller $\alpha$) in the QDQC operation.

Indeed, the error discussed here, which originates from the
interaction between the double-dot and its external control,
is relevant for all the other external ``knobs'' that are
used to control the evolution of the double-dot states.
To design a practical QDQC, one has to identify all the
possible external noise sources and tunes the system
parameters accordingly so as to prevent these noises from
overwhelming the operations of the QDQC.

\subsection{Future directions}

In the current manuscript we studied in detail the Hilbert
space structure for a two-electron two-dot 
artificial hydrogen molecule situation.  
It is important to emphasize that 
detailed theoretical calculations of the type carried out 
in this paper will be critical in determining the feasibility
and the practicality of all the proposed semiconductor-based
solid state QC architectures \cite{LD,Steel,BKane1}, not just
the spin-based QDQC considered in our work.  Given this crucial
importance of theory in providing the QC architectural basis
it is quite surprising that no such detailed calculations 
have earlier been reported in the literature in spite of
very extensive research activity in the subject of QC.
Indeed, there are many other theoretical questions that need to
be answered for the quantum dot quantum computer architecture.
For example, an accurate description
of the confinement potential is an important ingredient
of a quantum computer, as quantum computation requires an exact
knowledge of the qubit wavefunctions.  In addition, currently
there is no systematic calculation of spin relaxation in GaAs 
quantum dots, which will clearly be needed for a better
understanding of spin coherence issues.

As for further improving the calculation of electron
exchange coupling in the two-dot configuration, the main problem
is to obtain a more accurate description of electron correlations.
In the approaches we used in the current paper, the UHF method
self-consistently evaluated single particle wavefunctions, but
only a single Slater determinant is used as the two-electron 
wavefunction.  No two-electron correlation is included.  On the
other hand, the molecular orbital method uses a small number of
rigid single-electron wavefunctions (harmonic well single particle
orbitals), but many
two-electron orbitals are included to minimize the energy of the
system.  To improve upon the results obtained here, a self-consistent
calculation with CI is needed.  Namely, one can solve the 
Hartree-Fock equations self-consistently, then use these
HF wavefunctions as an improved basis
to form a number of Slater determinants (instead of just one as
in the HF calculation).  The two-electron problem can then be solved 
on the basis of these Slater determinants. Generally, the larger 
the basis the more accurate is the result.  Furthermore,
a linear-combination-of-atomic-orbital (LCAO) approach can be used
to partially solve the dilemma in the choice of gauge.
Such CI calculations, which we are currently pursuing, are
notoriously computationally demanding, and real progress
toward truly realistic calculations is expected to be slow.

As it is very difficult to precisely trap a single electron in
each quantum dot, one can consider using multi-electron quantum
dots as effective spin-$\frac{1}{2}$ qubit.  Thus an important
problem would be to study a multi-electron two-dot system,
in other words, a quantum dot Na$_2$ (or Cl$_2$, or others) 
molecule instead of H$_2$.  The objective of such a calculation is
two-fold. Firstly, at certain fillings there might exist effective
spin-$\frac{1}{2}$ states for a multi-electron system, so that the
`single electron' quantum dot requirement in the current proposal 
can be relaxed.  Secondly, such a calculation is also relevant in 
the general study of quantum dots.  We are currently pursuing
such calculations as well.

\section{Conclusion}

We have studied a quantum dot hydrogen molecule as the
basic elementary gate for a quantum computer based on 
electron spins in quantum dots.  By using both Hartree-Fock
approximation and a molecular orbital theory we determine the
excitation spectrum of two electrons in two horizontally coupled
quantum dots, and study its dependence on an external magnetic
field.  We particularly focus on the splitting of the lowest
singlet and triplet states---the exchange coupling, the double 
occupation probability of the lowest states, and the relative 
energy scales of these states.  We find that in our 
chosen configuration and for reasonable GaAs dot-based parameters
the exchange coupling has a maximum of about $0.2$ to $1.1$ meV at
zero magnetic field as we vary the central barrier height
from about 9.61 meV to 3.38 meV when the dots are separated
by 30 nm.  When the inter-dot separation increases to 40 nm,
the exchange coupling decreases to below 0.3 meV.
There exists a singlet-triplet crossing for all the cases
for an applied magnetic field of about
$4$ Tesla, above
which the triplet state becomes the ground state of the
two-electron system.  At zero magnetic field,
the double occupation probability in
the ground singlet state is found to be as large as 22\%
with a 3.38 meV central barrier when the two dots are 
separated by 30nm, and as small as 1.7\% with a 
11.03 meV central barrier
when the inter-dot distance is 40 nm.  Both the exchange
coupling and the double occupation probability have 
similar dependence on the inter-dot distance and the
central barrier height at zero magnetic field, so that
it is difficult to have a configuration with large exchange
coupling and vanishing double occupation probability (which
means a vanishingly small error rate).  At finite magnetic
field, on the other hand, it is possible to have a finite
(albeit negative) exchange coupling and a small double 
occupation probability simultaneously.
We discuss in detail the necessary
conditions for the validity of the effective mass
envelope function approach, finding this approximation to 
be valid for our problem.
We also discuss the applicability of various quantum
chemical approaches in the current context
of quantum dot quantum computation
in dealing with few-electron problems,
such as the Hartree-Fock self-consistent field method, 
the molecular
orbital method, the Heisenberg model, and the Hubbard model.
In particular, we point out that configuration interaction
calculation is needed for any realistic
description of electron wavefunctions.  The difference
between the non-product form of a Slater determinant and
a truly entangled state is discussed.  The presence of
singlet-triplet crossing in a coupled dot system is also
studied.  In addition, we discuss various important issues in
quantum dot quantum computing, such as controls needed, 
spin decoherence channels in semiconductors, adiabatic
transitions, and errors in spin evolution.
Our results should form a reasonable semi-realistic basis for 
discussing spin-based quantum dot quantum computer
architectures, and should also be useful for various
studies of quantum dot artificial molecule systems.

This work is supported by the Laboratory for Physical Sciences
(LPS) at the University of Maryland, 
the US-ONR, and DARPA.  We would also like to thank 
useful conversations with B. Kane (particularly on the discussion
of external noise due to gate voltage), D. Loss, G. Burkard,
D. DiVincenzo, and J. Fabian.

\newpage
\begin{figure}
\centerline{
\epsfxsize=3.5in
\epsfbox{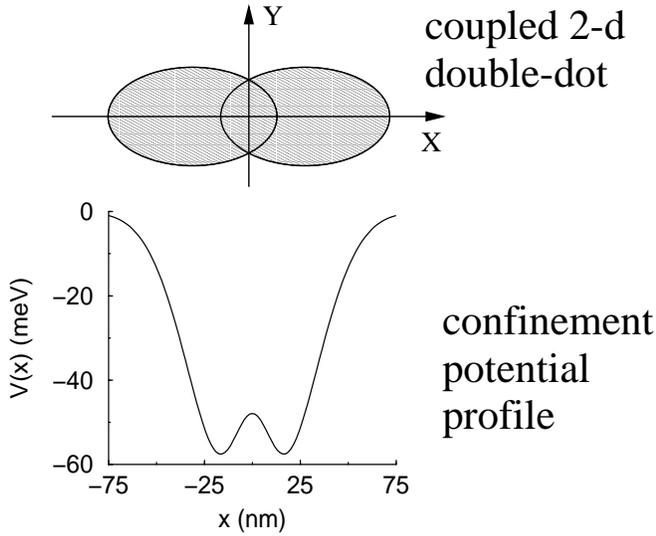}}
\vspace*{0.1in}
\protect\caption[Schematics of a coupled dot system]
{\sloppy{
This is a schematics of the double dot system we studied.
We use Gaussian potential wells and a Gaussian central barrier.  
Unless otherwise specified, the
dot size is 30 nm in radius.  
When the two dots are separated by 30 nm, we study
3 cases where the central
potential barrier has strength $V_b$ of 20, 25, and 30 meV, 
corresponding to effective barrier heights of 3.38, 6.28, and 
9.61 meV respectively.  When the two dots are separated
by 40 nm, we show results of 3 cases where $V_b$ takes the
values of 13.86, 18.17, and 20 meV, corresponding to 
actual barrier heights of 6.28, 9.61, and 11.03 meV.
}
}
\label{fig1}
\end{figure}
\begin{figure}
\centerline{
\epsfxsize=3.5in
\epsfbox{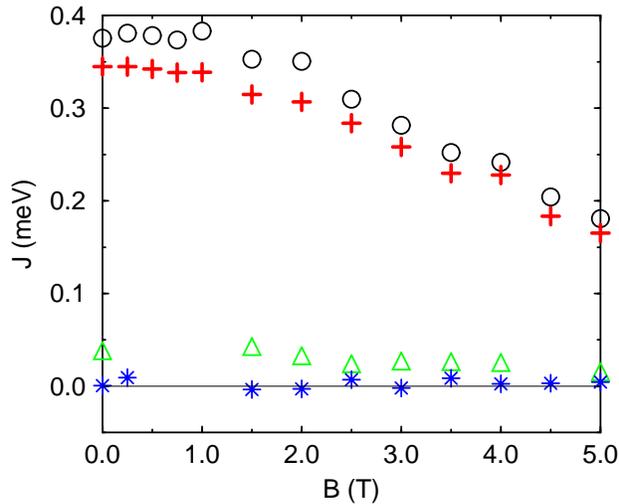}}
\vspace*{0.1in}
\protect\caption[Energy splitting between parallel and opposite
spin states in UHF, as a function of magnetic field]
{\sloppy{
Here we show the magnetic field (B) dependence of
the energy splitting (J) between parallel and opposite
spin states calculated by unrestricted Hartree-Fock approach.
The two higher energy curves are for dots with 30 nm radius, 
30 nm inter-dot separation, and 20 meV $V_b$.  
The lower energy ones are for dots with 70 nm radius and 80 nm
dot separation.  Between the two higher energy sets of data, 
the slightly lower one has a slightly thicker ($l_{bx}$ larger
by 2 nm) central
barrier.  The two sets of data for larger dots differ by 
central barrier heights ($V_b$) of 20 and 40 meV.
}
}
\label{fig2}
\end{figure}
\begin{figure}
\centerline{
\epsfxsize=3.5in
\epsfbox{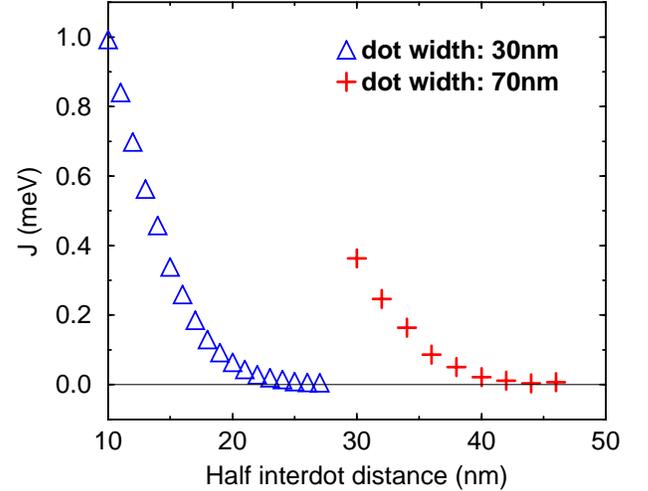}}
\vspace*{0.1in}
\protect\caption[Energy splitting between parallel and opposite
spin states in UHF, as a function of inter-dot distance]
{\sloppy{
Here we show the inter-dot distance dependence of
the energy splitting (J) between parallel and opposite
spin states calculated by unrestricted Hartree-Fock approach.
The left set of data corresponds to the small dot (30 nm radius
and 20 meV $V_b$) 
case while the right set of data to the large dot (70 nm radius
and 20 meV $V_b$) 
case.  Steep decrease in the energy splitting is present in both 
cases.
}
}
\label{fig3}
\end{figure}
\begin{figure}
\centerline{
\epsfxsize=3.5in
\epsfbox{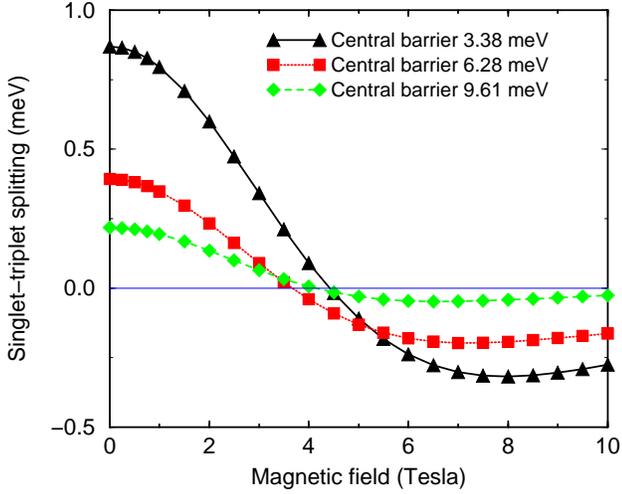}}
\vspace*{0.1in}
\protect\caption[Singlet-triplet splitting in a Hund-Mulliken
calculation]
{\sloppy{
Here we show the magnetic field dependence of the
singlet-triplet splitting in a Hund-Mulliken 
calculation for two electrons in a double dot
(of 30 nm radii) separated by 30 nm.  
Results of three different barrier heights
are shown.  The exchange coupling depends 
sensitively on the central barrier height.
}
}
\label{fig4}
\end{figure}
\begin{figure}
\centerline{
\epsfxsize=3.5in
\epsfbox{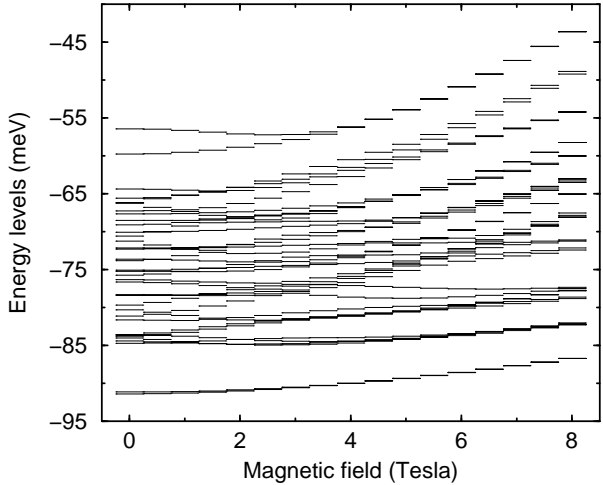}}
\vspace*{0.1in}
\protect\caption[Energy spectra of a molecular orbital
calculation involving both s and p orbitals, as functions
of magnetic field]
{\sloppy{
Here we show the magnetic field dependence of the
energy spectra in a molecular orbital 
calculation where both s and p single-electron 
orbitals are used.  The inter-dot distance is 30 nm,
and the central barrier $V_b$ is 30 meV, corresponding 
to an actual barrier height of 9.61 meV.
}
}
\label{fig5}
\end{figure}
\begin{figure}
\centerline{
\epsfxsize=3.5in
\epsfbox{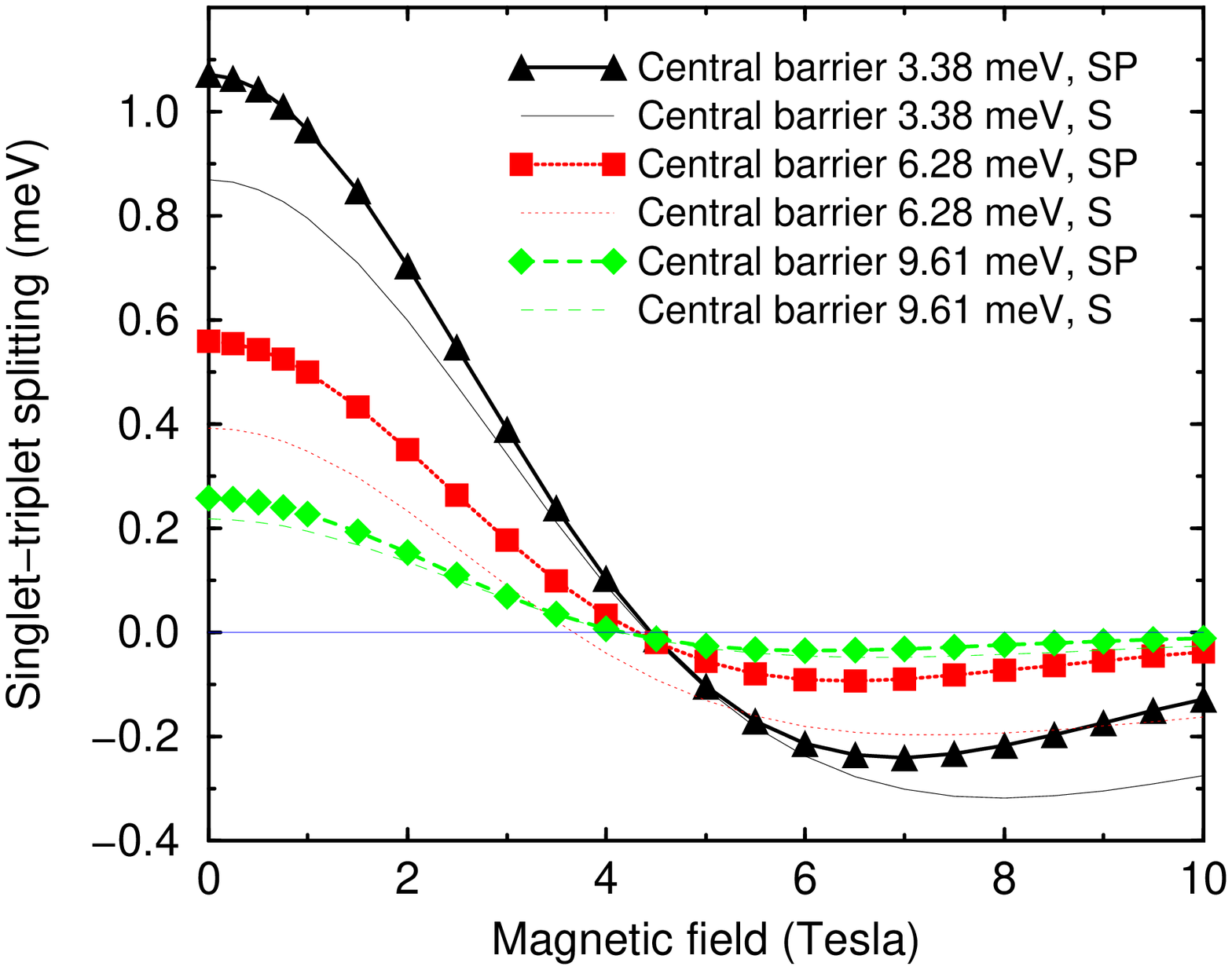}}
\vspace*{0.1in}
\protect\caption[Exchange coupling (splitting between the
singlet and triplet states) as a function of magnetic field]
{\sloppy{
Here we show the magnetic field dependence of the
exchange coupling in a molecular orbital 
calculation with both s and p 
single-electron orbitals. 
The inter-dot distance is 30 nm.
Results of three different barrier heights
are shown, together with the results (in thin lines)
from the Hund-Mulliken calculation for comparison.  The
exchange couplings from the full calculation are
about 20\% larger at zero magnetic field than those 
obtained from the Hund-Mulliken calculation.
}
}
\label{fig6}
\end{figure}
\begin{figure}
\centerline{
\epsfxsize=3.5in
\epsfbox{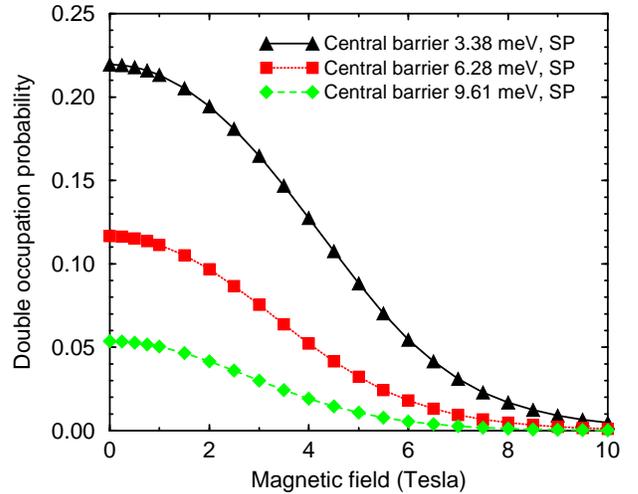}}
\vspace*{0.1in}
\protect\caption[Double occupation probability as a function 
of magnetic field] 
{\sloppy{
Here we show the magnetic field dependence of the
double occupation probability in a molecular orbital 
calculation with both s and p single-electron orbitals. 
This probability characterizes 
the double occupation occurring in the 
single electron ground state of the left 
dot.  It is clear that the two lower barrier cases, with
their large double occupation probabilities, are not appropriate
for the purpose of quantum computing at small magnetic fields.
}
}
\label{fig7}
\end{figure}
\begin{figure}
\centerline{
\epsfxsize=3.5in
\epsfbox{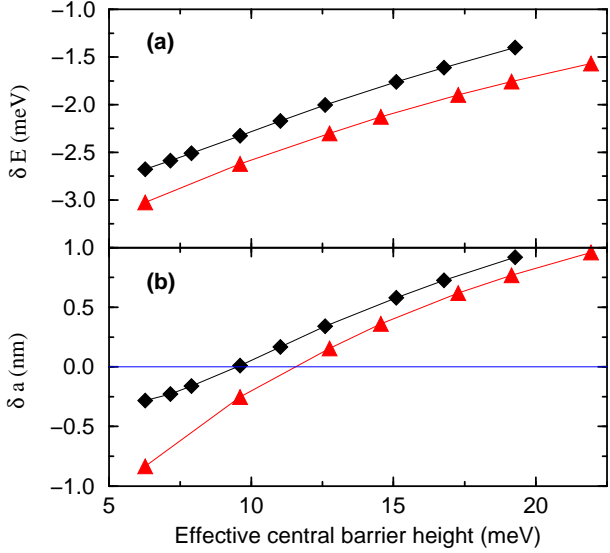}}
\vspace*{0.1in}
\protect\caption[Variational parameters as a function of 
effective central barrier height]
{\sloppy{
Here we show the central barrier height dependence of the
variational parameters in our study when 
the inter-dot distance is 40 nm.  Panel (a) shows the change
$\delta E$ in the fitting well parabolicity, 
while panel (b) shows the change $\delta a$ in
the locations of the two fitting wells (symmetric about
the origin.
The decrease in parabolicity and inter-orbital distance
indicates an analogy to orbital contraction and bonding
in molecular physics.
}
}
\label{fig8}
\end{figure}
\begin{figure}
\centerline{
\epsfxsize=3.5in
\epsfbox{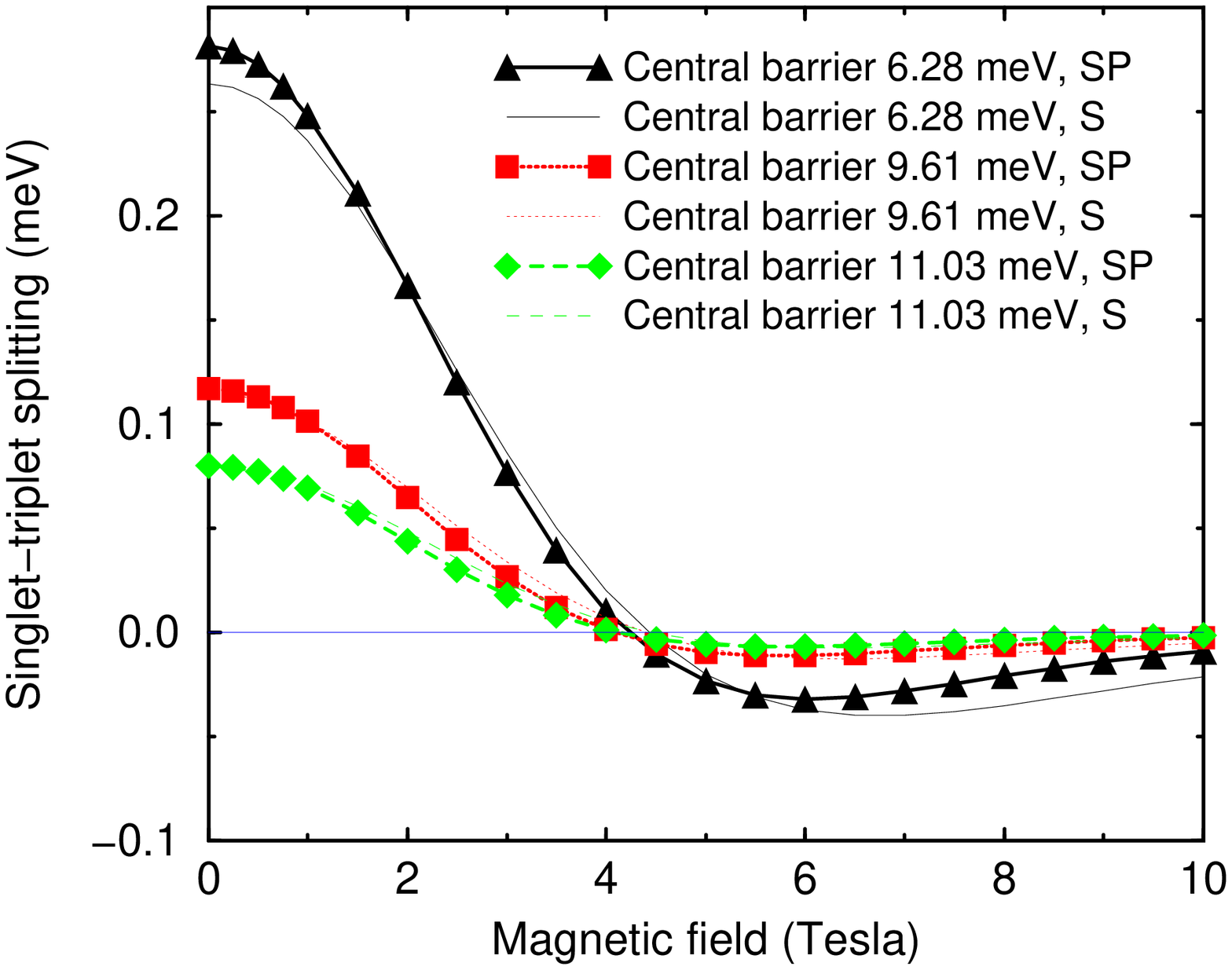}}
\vspace*{0.1in}
\protect\caption[Exchange coupling (splitting between the
singlet and triplet states) as a function of magnetic field]
{\sloppy{
Here we show the magnetic field dependence of the
exchange coupling in a molecular orbital 
calculation using both s and p 
single-electron orbitals. 
The inter-dot distance is 40 nm.
Results of three different barrier heights
are shown, together with the results (in thin lines)
from the Hund-Mulliken calculation for comparison.  The
exchange couplings from the full calculation are
only slightly larger at zero magnetic field than those 
obtained from the Hund-Mulliken calculation, but there
are some differences at finite magnetic fields.
}
}
\label{fig9}
\end{figure}
\begin{figure}
\centerline{
\epsfxsize=3.5in
\epsfbox{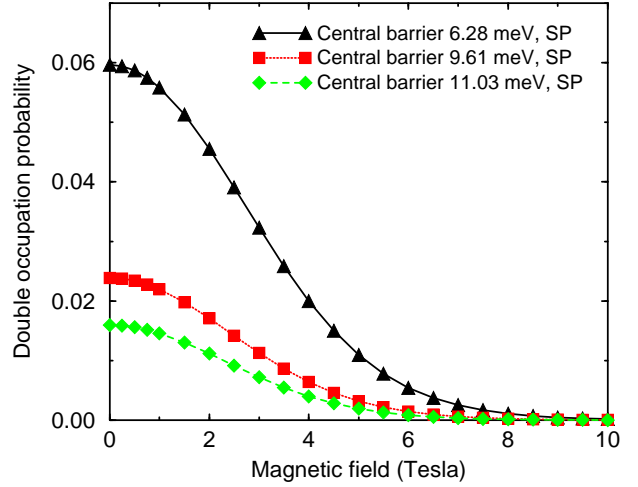}}
\vspace*{0.1in}
\protect\caption[Double occupation probability as a function 
of magnetic field] 
{\sloppy{
Here we show the magnetic field dependence of the
double occupation probability in a molecular orbital 
calculation with both s and p single-electron orbitals. 
The inter-dot distance is 40 nm.
This probability characterizes 
the double occupation occurring in the 
single electron ground state of the left 
dot.  At high magnetic fields the double occupation 
probabilities are vanishingly small for all three cases.
}
}
\label{fig10}
\end{figure}
\begin{figure}
\centerline{
\epsfxsize=3.5in
\epsfbox{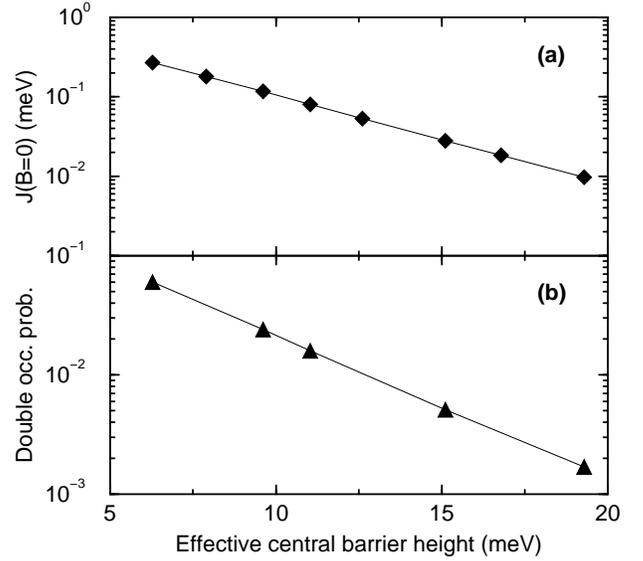}}
\vspace*{0.1in}
\protect\caption[Exchange coupling and double occupation 
probability as functions of the central barrier height
(at zero magnetic field)] 
{\sloppy{
Here we show the central barrier height dependence of the
exchange coupling $J$ and the
double occupation probability at zero magnetic field.  
The inter-dot distance here is 40 nm.
Both quantities decrease exponentially as 
functions of the central
barrier height.  The rates of these decreases
for both quantities are about the same.  As the central
barrier height varies in the shown range, $J$ changes 
from 0.27 meV to 0.0097 meV, while the double occupation
probability changes from 6\% to 0.17\%.
}
}
\label{fig11}
\end{figure}
\newpage

\begin{table}[h]
\begin{tabular}[h]{|l|c|c|c|}
\hline 
central potential barrier $V_b$ (meV) & 20 & 25 & 30 
\\ \hline
actual central barrier height (meV) & 3.38 & 6.28 & 9.61 
\\ \hline
change in parabolicity (meV) & 
-2.8281 & -2.3915 & -2.0044 \\ \hline
actual single particle excitation & 
8.4134 & 8.8499 & 9.2371 \\ 
energy at zero {\bf B} field (meV) & & & \\ \hline
change in fitting well location (nm) &
-0.2243 & -0.3779 & -0.1632 \\ \hline
actual fitting well location &
12.6343 & 14.2441 & 16.0822 \\ 
at zero {\bf B} field (nm) & & & \\ \hline
\end{tabular}
\caption[]{Here we tabulate the variational parameters for the
three different central barrier heights at 30 nm 
inter-dot distance.  The fitting well refers
to the isotropic parabolic wells we use to fit the two 
Gaussian wells.  We obtain the base parabolicity from the
second derivative at the bottom of the confinement potential
wells, and the base locations are the actual minima of the
confinement potential wells.}
\label{table1}
\end{table}
%

\end{document}